\documentclass[prd,nofootinbib,twocolumn,superscriptaddress,preprintnumbers]{revtex4-1}

\usepackage{placeins}
\usepackage{soul}
\usepackage{chngcntr}
\usepackage{color}
\usepackage{epsfig}  
\usepackage{graphicx}
\usepackage{tabularx}
\usepackage{xspace}
\usepackage{float}

\usepackage{color}
\definecolor{magenta}{rgb}{0.55, 0.0, 0.55}
\definecolor{orange}{rgb}{0.75, 0.25, 0.05}
\usepackage{hyperref}
\hypersetup{
     colorlinks   = true,
     citecolor    = magenta,
	 linkcolor=blue
}

\usepackage{verbatim}
\usepackage{amsmath}
\usepackage{relsize}
\usepackage{amssymb}
\usepackage{url}
\usepackage{bbold}
\usepackage{slashed}
\usepackage{array}

\usepackage{multirow}
\usepackage{threeparttable}
\usepackage{paralist}
\usepackage{xspace}
\usepackage{upgreek}
\usepackage{lipsum}

\newcommand{\bra}[1]{\left\langle #1 \right|}
\newcommand{\ket}[1]{\left|#1\right\rangle}

\newcommand\colvec[3][]{\begin{pmatrix}\ifx\relax#1\relax\else#1\\\fi#2\\#3\end{pmatrix}}

\newcommand{\beq}{\begin{equation}}
\newcommand{\beqn}{\begin{eqnarray}}
\newcommand{\eeq}{\end{equation}}
\newcommand{\eeqn}{\end{eqnarray}}

\newcommand{\eV}{ \textrm{eV} }

\newcommand{\MeV}{ \textrm{MeV} }
\newcommand{\GeV}{ \textrm{GeV} }

\newcommand{\bfR}{ {\bf R} }
\newcommand{\bfE}{ {\bf E} }

\newcommand{\bfK}{ {\bf K} }
\newcommand{\bfL}{ {\bf L} }
\newcommand{\bfv}{ {\bf v} }
\newcommand{\bfr}{ {\bf r} }
\newcommand{\bfk}{ {\bf k} }
\newcommand{\bfp}{ {\bf p} }
\newcommand{\bfq}{ {\bf q} }
\newcommand{\bfe}{ {\bf e} }

\newcommand{\Zion}{ Z_{\rm ion} }

\newcommand{\ourtitle}{The Migdal effect in semiconductors}

\graphicspath{{./figures/}}

\def\mysection#1{{\bf #1 ---} }

\definecolor{cerulean}{rgb}{0., 0.42,0.9}

\begin{document}

\title{\ourtitle}
\author{Simon Knapen}
\email{simon.knapen@cern.ch}
\affiliation{CERN, Theoretical Physics Department, Geneva, Switzerland}
\author{Jonathan Kozaczuk}
\email{jkozaczuk@physics.ucsd.edu}
\affiliation{Department of Physics, University of California, San Diego, CA 92093, USA}
\author{Tongyan Lin}
\email{tongyan@physics.ucsd.edu}
\affiliation{Department of Physics, University of California, San Diego, CA 92093, USA}

\begin{abstract}\noindent
When a nucleus in an atom undergoes a collision, there is a small probability to inelastically excite an electron as a result of the Migdal effect. In this Letter, we present a first complete derivation of the Migdal effect from dark matter--nucleus scattering in semiconductors, which also accounts for multiphonon production. The rate can be expressed in terms of the energy loss function of the material, which we calculate with density functional theory (DFT) methods. Because of the smaller gap for electron excitations, we find that the rate for the Migdal effect is much higher in semiconductors than in atomic targets. Accounting for the Migdal effect in semiconductors can therefore significantly improve the sensitivity of experiments such as DAMIC, SENSEI and SuperCDMS to sub-GeV dark matter. 
\end{abstract}

\maketitle

\mysection{Introduction} 
Direct detection of nuclear recoils from sub-GeV dark matter (DM) is challenging because the typical energy deposited in an elastic nuclear recoil scales as $E_N\sim  m_\chi^2 v_\chi^2/m_N$, which is exceedingly difficult to detect for lighter dark matter candidates. Furthermore, low-energy nuclear recoils do not deposit much energy in the form of charge (electrons) or scintillation light, which are primary detection channels in many experiments. These considerations are drivers for new technologies and experiments capable of lower thresholds and phonon-based detection~\cite{Battaglieri:2017aum,DarkMatterBRN}.

In the meantime, existing experiments can significantly extend their reach by searching for inelastic scattering processes, during which additional excitations are created, such as ionization electrons~\cite{Ibe:2017yqa}, photons~\cite{Kouvaris:2016afs}, or plasmons~\cite{Kurinsky:2020dpb,Kozaczuk:2020uzb}. While there is generally a rate penalty for such processes, there are two key advantages: the unfavorable energy relation in the previous paragraph can be broken and the additional excitation gives rise to charge signals that are more readily detected. 

The Migdal effect  \cite{Migdal1939,Migdal:1977bq} describes inelastic collisions where atoms are excited or ionized during the initial, hard nuclear recoil. For $m_\chi\gtrsim$ 70 MeV, this process takes place on time scales much shorter than the time for the recoiling nucleus to travel a distance comparable to the interatomic spacing. It can therefore be factorized from secondary ionizations that can occur as the recoiling nucleus interacts with surrounding atoms, as described by Lindhard's theory \cite{Lindhard:1963aaa}. As we will show, in crystals the underlying physics of the primary (Migdal) and secondary (Lindhard) ionizations is closely related, but they take place on different time scales.

So far, the Migdal effect and other inelastic processes have primarily been studied in atomic targets. As applied to dark matter direct detection, the most complete derivation of the Migdal effect can be found in Ref.~\cite{Ibe:2017yqa}, while additional discussion can be found in Refs.~\cite{Bernabei:2007jz,Bell:2019egg,Baxter:2019pnz,GrillidiCortona:2020owp,Nakamura:2020kex,Liang:2019nnx,Liu:2020pat,Essig:2019xkx,Dolan:2017xbu}. The effect has been applied to set strong experimental limits in noble liquid targets~\cite{Akerib:2018hck,Aprile:2019jmx,Dolan:2017xbu,Essig:2019xkx}. Charge thresholds in low-threshold semiconductor experiments are even lower~\cite{Aguilar-Arevalo:2019wdi,Barak:2020fql,Amaral:2020ryn}, which makes them even better suited to exploit the Migdal effect. However, inelastic processes are less well-studied in crystal targets, in part due to the more complicated spectrum of excitations. Moreover, the existing calculation for atomic targets \cite{Ibe:2017yqa} relies on boosting the system to the rest frame of the recoiling nucleus. While convenient for atomic calculations, this method breaks down for semiconductors, as the rest frame of the crystal is a preferred frame.  So far, Refs.~\cite{Essig:2019xkx,Liang:2019nnx,Liu:2020pat} presented estimates of the Migdal effect in solid state targets, but as of now a fully robust, first principles derivation is still lacking. 

In this Letter, we show that the Migdal effect in semiconductors can be described by the $2 \to 3$ process of DM--nucleus scattering in association with a nucleus--electron Coulomb interaction. Furthermore, the details of the electronic band structure can be packaged into the well-studied energy loss function, ${\rm Im}(-1/\epsilon)$, where $\epsilon$ is the dielectric function of the material. When the energy deposited into electronic excitations is close to the plasma frequency, plasmons are resonantly excited. (A plasmon resonance can be thought of as a collective electron excitation or as a longitudinal in-medium photon mode.) A first calculation of the rate for inelastic DM--nucleus scattering with associated plasmon production was presented in Ref.~\cite{Kozaczuk:2020uzb}.  By directly calculating the non-resonant contributions, we show that plasmon production is simply the resonant component of the Migdal effect.

We will begin with a general description of the process in semiconductors and lay out the assumptions in our calculation. We  present our main results of the Migdal rate in semiconductors, and then clarify its relation to the atomic Migdal effect. Finally, we present sensitivity estimates for experiments using Si and Ge targets. We leave the bulk of our calculations for the Appendices.

\mysection{Description of process} 
For an elastic recoil off a free nucleus of mass $m_N$, the typical momentum deposited by sub-GeV DM is $q_N \approx m_\chi v_\chi \approx \MeV \times (m_\chi/\GeV)$ and the typical recoil energy is $E_N \approx m_\chi^2 v_\chi^2/m_N \approx 35\, \eV \times (m_\chi/\GeV)^2$, taking the example of a Si target. For sub-GeV DM, the energy and momentum scales are then comparable to various scales inherent to the crystal. Some care is therefore needed with respect to the regime of validity of our approximations. 

\begin{figure}[t]
\centering
\includegraphics[width=0.46\textwidth]{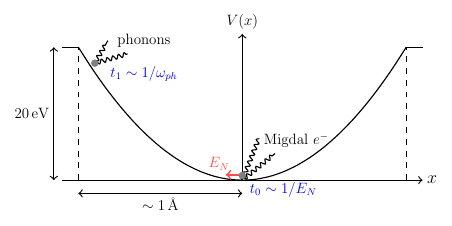}
\caption{We compute DM scattering off a nucleus in a harmonic crystal using the impulse approximation, which is valid when the time scale of the initial collision $(t_0)$ is short compared to the time scale to traverse its potential well $(t_1)$, set by the phonon frequency $1/\bar\omega$. This holds for $m_\chi \gtrsim 70$ MeV.
\label{fig:impulseapprox}}
\end{figure}

Concretely, in a typical crystal, each nucleus sits in an approximately harmonic potential with size of $\sim$\AA\ and frequency $\bar\omega \approx 30-50$ meV (see Fig.~\ref{fig:impulseapprox}). We will deal with $m_{\chi} \gg~$10 MeV such that the  inverse momentum transfer $1/q_N \ll\, $\AA. Then we can consider the interaction of the DM with a single nucleus, the so-called \emph{incoherent approximation}\footnote{Coherent scattering with multiple nuclei (that is, phonon excitation) occurs for $m_\chi \lesssim \MeV$ and was treated in Refs.~\cite{Knapen:2017ekk,Griffin:2018bjn,Griffin:2019mvc,Trickle:2019nya,Cox:2019cod,Campbell-Deem:2019hdx,Griffin:2020lgd}.}. We will thus compute DM scattering off a nucleus in the ground state of the potential, with associated nucleus--electron interaction.

To treat the excited states of the nucleus, we will rely on the \emph{impulse approximation}, which is valid if the collision happens quickly relative to the time scale set by the potential well, $1/\bar\omega$. (See e.g.~\cite{watson1996neutron}) The initial DM-nucleus collision and the emission of the Migdal electrons take place on a time scale $\sim 1/E_N$, during which the nucleus remains near the minimum of the potential well (see Fig.~\ref{fig:impulseapprox}). Only at a much later time $1/\bar\omega$, the nucleus reaches the edge of the unit cell and loses its residual kinetic energy to phonons, or becomes unbound, depending on the initial collision energy. The impulse approximation thus allows us to model the excited states as plane waves for the duration of the hard collision\footnote{In the atomic Migdal effect, the recoiling ion is often modeled as a semiclassical current. In this context, the ``impulse approximation'' corresponds to a stronger set of assumptions than those employed here.}.  We show this in Appendix~\ref{app:freeionapprox} by calculating the all orders, multiphonon response of a harmonic crystal and find that the approximation holds if the momentum transfer satisfies $q_N\gtrsim \sqrt{2m_N\bar\omega}$. For DM with the standard velocity profile, this implies $m_\chi \gtrsim 70$ MeV. The incoherent approximation is then satisfied automatically as well. In practice, we will apply a set of cuts to exclude regions of the phase space where the approximations start to break down.

In this discussion we have focused on the DM momentum transfer to the nucleus, $q_N$, because the momentum deposited in electrons will be much smaller. The Migdal rate then factorizes as: 
\begin{equation}
    \label{eq:sigma_factorized}
   \frac{d\sigma_{\rm ion}}{dE_N d\omega}\approx  \frac{d\sigma_{\rm el}}{dE_N} \frac{dP(E_N)}{d\omega}
\end{equation}
where $\omega$ is the energy deposited into electronic excitations, $d\sigma_{\rm el}/dE_N$ the elastic DM-nucleus cross section and $dP(E_N)/d\omega$ is the differential ionization probability.  This is identical to the expansion made in the bremsstrahlung of a soft photon from a heavy charged particle, and we refer to it as the \emph{soft limit}. The soft limit holds as long as $|\bfq_N\cdot \bfk|\ll m_N \omega$ and $k\ll q_N$. Estimating $q_N\sim v m_\chi$ and $k\sim 1-10$ keV, this translates to \mbox{$50\; \text{MeV} \lesssim m_\chi\lesssim 1$ GeV} and $\omega \gtrsim $ eV, which covers the most relevant parameter space.
While our formal result does not rely on the soft limit, it is a useful technical and conceptual simplification when performing the phase space integrals, and is valid whenever the impulse approximation holds. 

Finally, we will treat the nuclei and tightly-bound core electrons together as a particle with charge $\Zion$, and only consider excitations of the valence electrons. In other words, we assume an ion potential which behaves as $\Zion e/r$ for large $r$ compared to the wavefunctions of the inner shell electrons. To account for the effective ion charge at shorter distances, we include a momentum-dependent $\Zion(k)$ in the Fourier-transformed potential, which we obtain using tabulated ionic form factors~\cite{BrownXray}. While the electron-ion momentum exchange will be $\lesssim 10$ keV such that the long-range behavior of the potential is most important, including $\Zion(k)$ leads to $O(1)$ rate increases.

\mysection{Calculation} The Migdal rate is given by the leading order expansion in both the DM-nucleus and the electromagnetic interactions, analogous to bremsstrahlung. We assume a contact interaction between the DM and the nuclei, given by the Hamiltonian $H_\chi = (2 \pi b_\chi/m_\chi) \delta(\bfr_\chi-\bfr_N)$ for $m_\chi\ll m_N$, with $b_\chi$ the DM-nucleus scattering length and $\bfr_\chi, \bfr_N$ the position operators for the DM and nucleus. (The DM-nucleus elastic cross section is therefore given by $\sigma_N=4\pi b_\chi^2 = A^2 \sigma_n$, with $A$ and $\sigma_n$ respectively the atomic mass number and DM-nucleon elastic cross section.) The electron-nucleus interaction is $H_e =  \int\! d^3\bfr' \epsilon^{-1}(\bfr',\bfr,\omega) \, Z_{\text{ion}} \alpha/| \bfr' - {\bf r}_{N}|$ with $\bfr$ the position operator for the electron. $\alpha$ is the electromagnetic fine structure constant, and $\epsilon$ is the frequency-dependent, microscopic dielectric function, which  encodes the screening by the spectator valence electrons. 

In a crystal, the linear response depends on both $\bfr'$ and $\bfr$, up to the lattice periodicity. The Fourier transform of the response $\epsilon^{-1}(\bfr',\bfr,\omega)$ can then be written as $\epsilon^{-1}_{\bfK\bfK'}(\bfk, \omega)\equiv\epsilon^{-1}(\bfk+\bfK,\bfk+\bfK', \omega)$ where $\bfk$ is in the first Brillouin Zone (BZ) and $\bfK, \bfK'$ are reciprocal lattice vectors. $\epsilon^{-1}_{\bfK \bfK'}$ can be regarded as a matrix in the reciprocal lattice vectors, but for Si and Ge we find the contribution of the off-diagonal pieces to be subleading. Here we just present results assuming a diagonal response matrix $\epsilon^{-1}_{\bfK\bfK}$ and reserve the full result for Appendix~\ref{app:semi_fullderivation}. Including the momentum-dependent ion charge,  $H_e$ can then be written as \begin{equation}\label{eq:HeFouriermaintext}
    H_e = - 4\pi \alpha  \sum_{\bfK} \! \int\!\! \frac{d^3\bfk}{(2\pi)^3}  \frac{\Zion(|\bfk + \bfK|) \,  e^{i (\bfr - \bfr_N) \cdot (\bfk + \bfK)}}{\epsilon_{\bfK\bfK}(\bfk, \omega) \, |\bfk + \bfK|^2}.
\end{equation}

We can apply Fermi's golden rule with second-order perturbation theory to compute the cross section for DM--nucleus inelastic scattering. We take the initial ions to be in a ground state of a harmonic crystal potential. Following the impulse approximation, we use plane waves for intermediate and final states. 
Meanwhile, the electron states are treated as Bloch states. Though the computation itself is a straightforward application of second order perturbation theory, the formulas and derivation are somewhat lengthy due to the appearance of the reciprocal lattice vectors and a form factor for the recoiling ion. We refer the reader to Appendix~\ref{app:semi_fullderivation} for further details, and only present the final result here:
\begin{widetext}
\begin{align}
     \frac{d\sigma}{d\omega}& =\frac{2\pi^2  A^2 \sigma_n}{m_\chi^2 v_\chi} \!     \int\!\frac{d^3\bfq_N}{(2\pi)^3}   \int\!\!\frac{d^3\bfp_f}{(2\pi)^3} \, \delta(E_i-E_f-\omega-E_N) \times  F(\bfp_i-\bfp_f-\bfq_N-\bfk-\bfK)^2 \times \sum_{\bfK}\!  \int\!\!\frac{d^3\bfk}{(2\pi)^3} \frac{4 \alpha \Zion^2(|\bfk + \bfK|) }{|\epsilon_{\bfK\bfK}(\bfk,\omega)|^2}
     \nonumber \\
     &\times \Bigg[\frac{1}{\omega-\bfq_N\cdot (\bfk+\bfK)/m_N}-\frac{1}{\omega}\Bigg]^2  \underbrace{ \frac{4\pi^2\alpha}{V}\sum_{\bfp_e} \frac{| [\bfp_e+\bfk| e^{i\bfr\cdot\bfK}|\bfp_e]_\Omega |^2}{ |\bfk+\bfK|^2} \left(f(\bfp_e)-f(\bfp_e+\bfk)\right)\delta(\omega_{\bfp_e+\bfk}-\omega_{\bfp_e} -\omega) }_
     {\mathlarger{ \textrm{Im}\, [ \epsilon_{\bfK \bfK}(\bfk, \omega)] } } \label{eq:ratefinalmaintext}
\end{align}
\end{widetext}
where  $\bfq_N$ and $\bfp_f$ are the final ion and DM momentum, respectively, and $\bfk + \bfK$ is the momentum deposited to the electrons. In the first part of \eqref{eq:ratefinalmaintext}, we see the same factors and phase space integral that appear for elastic DM-nucleus scattering, except with the additional energy $\omega$ being deposited in electronic excitation. While for free nucleus scattering there would be a momentum conservation delta function, here we have a form factor $F$ which encodes the details of the ion ground state, and for a harmonic crystal it is given by
\begin{equation}
   F(\bfp_i-\bfp_f-\bfq)\equiv \left(\frac{4\pi}{m_N\bar\omega}\right)^{3/4}e^{\frac{-|\bfp_i-\bfp_f-\bfq|^2}{2m_N \bar\omega}}
\end{equation}
where $\bar \omega$ is an oscillator frequency, averaged with respect to the density of states $D(\omega)$ and the thermal Bose factor. The remaining pieces of \eqref{eq:ratefinalmaintext} contain the probability for exciting the electron. We sum over all initial and final electron states $\bfp_e$ and $\bfp_e + \bfk$, weighted by the occupation numbers $f$, and where band indices have been suppressed. The electronic wavefunction overlaps $[\bfp_e+\bfk| e^{i\bfr\cdot\bfK}|\bfp_e]_\Omega$ are performed over the unit cell, and $V$ is the volume of the crystal.

In \eqref{eq:ratefinalmaintext}, the bracketed quantity can be rewritten in terms of the imaginary part of the dielectric function in the random phase approximation, $\textrm{Im}\, [ \epsilon_{\bfK \bfK}(\bfk, \omega)]$. Then we can write  $\textrm{Im}\, [ \epsilon_{\bfK \bfK}(\bfk, \omega)]/|\epsilon_{\bfK \bfK}(\bfk, \omega)|^2 = \textrm{Im}\, [-1/ \epsilon_{\bfK \bfK}(\bfk, \omega)]$, which is the energy loss function (ELF) governing energy loss of charged particles in a material. Physically,  the ion-electron interaction in the inelastic process can be encapsulated in the same ELF as ions passing through a material. Since the ELF is a well-measured and calculated quantity in many materials, this provides a useful starting point for numerical evaluations of \eqref{eq:ratefinalmaintext}.

In the soft limit $|\bfk + \bfK| \ll |\bfq_N|$, the cross section factorizes as in \eqref{eq:sigma_factorized}, and the form factor $F$ only modifies the elastic recoil cross section. Then the differential ionization probability is 
\begin{align}
\label{eq:Psemi_compare}
\frac{dP}{d\omega}= & \, \frac{(4 \pi \alpha)^2}{\omega^4} \sum_{\bfp_e}\int\!\!\frac{d^3\bfk}{(2\pi)^3}  \Zion^2(k)  \frac{|\bfv_N\cdot \bfk|^2}{k^4} \frac{\left|[\bfp_e+\bfk| \bfp_e]_\Omega \right|^2}{V |\epsilon(\bfk,\omega)|^2 } \nonumber \\ & \times\left(f(\bfp_e)-f(\bfp_e+\bfk)\right)\delta(\omega_{\bfp_e+\bfk}-\omega_{\bfp_e} -\omega) \\
=& \, 4\alpha\int\!\!\frac{d^3\bfk}{(2\pi)^3} \frac{\Zion^2(k)}{\omega^4} \frac{ |\bfv_N\cdot \bfk|^2}{k^2} \text{Im}\left(\frac{-1}{\epsilon(\bfk,\omega) } \right).     \label{eq:Psemi_epsilon}
\end{align}
with $\bfv_N\equiv \bfq_N/m_N$. This simplified formula is only valid for $\bfk$ in the first Brillouin zone, see Appendix~\ref{app:semi_fullderivation} for the full expressions used in our numerical results.
Eq.~\eqref{eq:Psemi_epsilon} was also derived in \cite{Kozaczuk:2020uzb}, but that work did not account for the ion ground state or electron momentum transfers outside of the first BZ, since it was focused on long-wavelength plasmons. Furthermore, \cite{Kozaczuk:2020uzb} used an analytic approximation for $\epsilon(\bfk,\omega)$ near the plasmon pole. In the results below, we will study the impact of accounting for the ion ground state and use numerical calculations of $\epsilon(\bfk,\omega)$ valid away from the plasmon resonance. Before doing so, we clarify the relation of this process with the atomic Migdal effect.

\mysection{Comparison with atomic Migdal effect}
In Migdal's original calculation \cite{Migdal1939,Migdal:1977bq} for an atomic target, the ground state of the electron cloud ($\ket{i}$) is first boosted to the rest frame of the moving nucleus $\ket{i} \to e^{i m_e \bfv_N \cdot \sum_\beta \bfr_{\beta} }  \ket{i}$. He then computes the overlap with the excited states $\bra{f}$
\begin{align}\label{eq:M_boost}
\mathcal{M}_{if} = \bra{f} e^{i m_e \bfv_N \cdot \sum_\beta \bfr_\beta }  \ket{i}\approx i m_e\bra{f} \bfv_N \cdot \mathsmaller{\sum_\beta} \bfr_\beta   \ket{i}
\end{align}
where $\beta$ runs over all the electrons in the atom.  The transition probabilities $|{\cal M}_{if}|^2$ can then be evaluated with known atomic wave functions, and it was found that single ionizations dominate for sub-GeV dark matter~\cite{Ibe:2017yqa}.

To demonstrate the connection with the semiconductor Migdal effect derived above, we instead rewrite \eqref{eq:M_boost} using the following operator identity: $\langle f |  \sum_\beta \bfr_\beta | i \rangle  = -i\langle f| \sum_\beta \bfp_\beta | i \rangle/m_e \omega =  i \langle f | \sum_\beta [\bfp_\beta, H_{0}] | i \rangle /m_e\omega^2$,
where again $\omega = E_f - E_i$ is the total energy deposited and $H_0$ the electron Hamiltonian.
We assume a non-relativistic\footnote{Relativistic corrections can be important for inner shell electrons, but the rate is dominated by the non-relativistic outer shells.} Hamiltonian such that the $H_0$ is a sum of kinetic terms, Coulomb interaction terms between electrons, and the Coulomb interaction of the electrons with the nucleus. Then the commutator $\sum_\beta [\bfp_\beta, H_{0}]$ will be proportional to the total force from the nucleus, since the electron-electron forces cancel out. Contracting the matrix element with $\bfv_N$, we find (see also \cite{ROSEL198243})
\begin{align}
  \langle f | \mathsmaller{\sum_\beta}  \bfv_N \cdot \bfr_\beta | i \rangle = \frac{1}{m_e \omega^2}  \langle f|\mathsmaller{\sum_\beta} \frac{ Z_N \alpha \bfv_N \cdot \hat \bfr_\beta}{|\bfr_\beta - \bfr_N|^2} | i \rangle.
   \label{eq:dipoleforce2}
\end{align}
with $\bfr_N$ the position operator of the nucleus.
In the right-hand-side matrix element above, we see the time derivative of the dipole potential from a nucleus which has been displaced by $|\bfr_N| \ll |\bfr_\beta|$. This is already very similar to the Coulomb potential in \eqref{eq:HeFouriermaintext} and suggestive of the same physical interpretation as in the semiconductor case. One can now take the Fourier transform and evaluate the transition probability for single ionizations:
\begin{align}
    \frac{dP(E_N)}{d\omega}  \approx& \left(\frac{4 \pi Z_N \alpha}{\omega^2} \right)^2 \sum_{i,f}    \Bigg|  \int\!\! \frac{d^3 \bfk}{(2\pi)^3} \frac{\bfv_N \cdot \bfk}{k^2} \, \langle f| e^{i \bfk \cdot \bfr} | i \rangle \Bigg|^2\nonumber\\& \times  \delta \left(E_i + \omega - E_f  \right) .
    \label{eq:atomic_fourier}
\end{align}
where we have dropped the $e^{-i \bfk \cdot \bfr_N}$ phase factor in the soft limit. We have thus shown that the atomic Migdal effect has a form nearly identical to \eqref{eq:Psemi_compare} for semiconductors, up to a few differences reflecting the different physical systems. In \eqref{eq:Psemi_compare}, the integral over $\bfk$ appears {\emph{outside}} the amplitude squared; this reflects conservation of crystal momentum in the semiconductor, which requires that the final state have momentum $\bfp_e + \bfk$, whereas the atomic states are not momentum eigenstates. The nucleus charge $Z_N$ appears here, since we are considering the all-electron wavefunctions, whereas in the semiconductor case we effectively integrated out the core electrons and treated the ion with effective charge $\Zion(k)$. Finally, in \eqref{eq:Psemi_compare}, we accounted for the in-medium dielectric screening $1/|\epsilon(\bfk, \omega)|^2$ due to all the other electrons, which is neglected in the atomic case.

With this result, we find an equivalent formulation of the atomic Migdal effect, \eqref{eq:atomic_fourier}, but which has a physical interpretation that applies also in semiconductors. Previously, Ref.~\cite{Essig:2019kfe} assumed the atomic formulation in \eqref{eq:M_boost} could be generalized directly to semiconductors. However, the boosting argument used to obtain  \eqref{eq:M_boost} does not apply in this case, since in a crystal there is a preferred coordinate frame. In other words, applying a boost operator would boost \emph{all} nuclei in the lattice. Specifically, the two approaches are not equivalent because the operator relation \eqref{eq:dipoleforce2} only applies for an individual atom: in the presence of a lattice of nuclei, we would have contributions from \emph{all} nuclei on the right-hand-side of \eqref{eq:dipoleforce2}.  Ref.~\cite{Liang:2019nnx} attempted to address this subtlety by using atom-centered localized Wannier functions in \eqref{eq:M_boost}, to mitigate the contribution from the remaining nuclei in the crystal. For Si and Ge in particular, the Wannier approach is however found to be computationally challenging due to slow numerical convergence \cite{Liang:2019nnx}. 

Between these different starting points, \eqref{eq:atomic_fourier} has a clear physical interpretation as the in-medium analogue of bremsstrahlung, which nicely generalizes to the semiconductor case. This interpretation is discussed further in Appendix~\ref{sec:semiclassical_atomic}. The final result in semiconductors can moreover be expressed in terms of the ELF, which can be calculated with a number of public codes. We therefore argue for this approach in generalizing the atomic Migdal effect.

\begin{figure*}[t]
\centering
\includegraphics[width=0.46\textwidth]{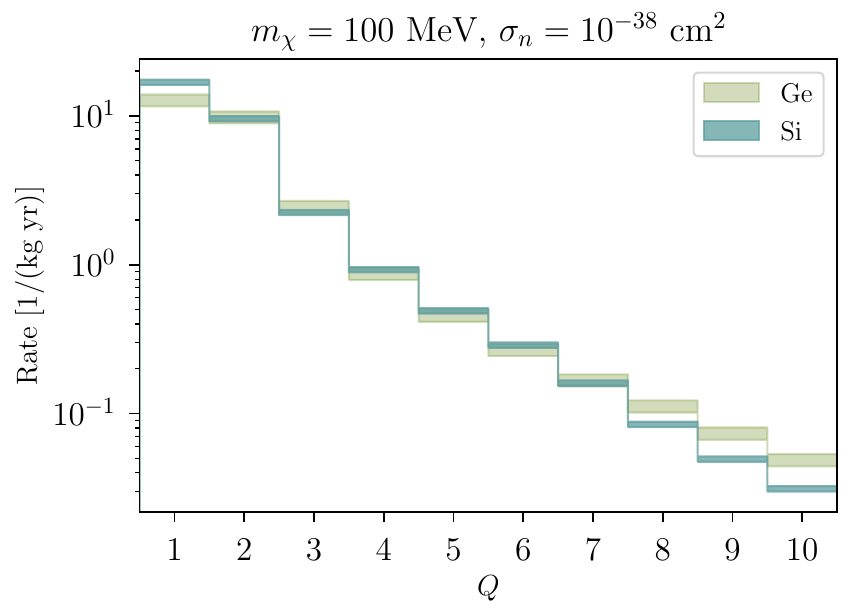}\hfill
\includegraphics[width=0.47\textwidth]{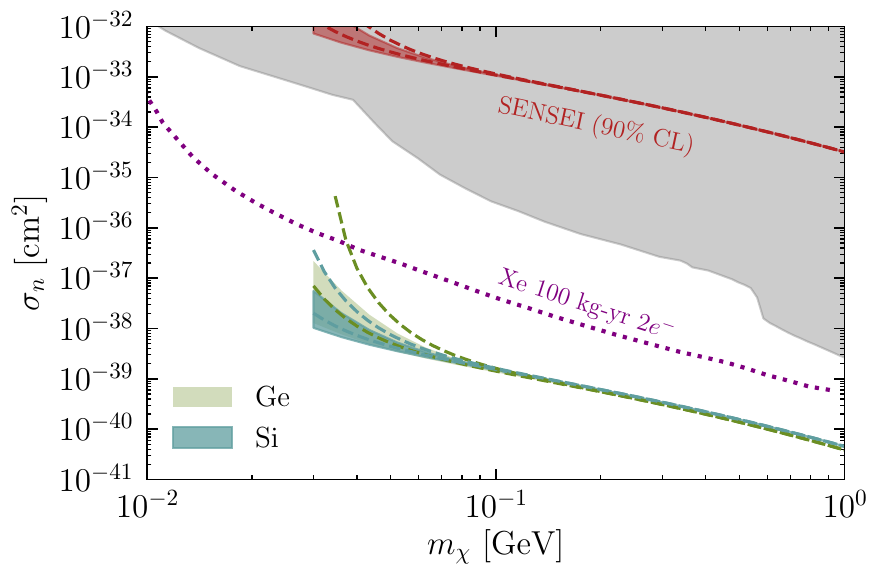}
\caption{({\bf Left}) Electron recoil spectrum in Si and Ge from the Migdal effect in semiconductors, assuming DM--nucleon cross section of $10^{-38}$ cm$^2$ and DM mass 100 MeV. The differential rate is translated into total number of electron-hole pairs created ($Q$) using Eq.~5.1~of Ref.~\cite{Essig:2015cda}.  ({\bf Right}) Expected 90\% CL sensitivity to DM--nucleon cross section $\sigma_n$ assuming a heavy mediator and 1 kg-year of exposure. We take $Q \ge 2$ and zero background, corresponding to an upper limit of 2.4 events.  The red line is a 90\% CL limit obtained using the recent upper limit on the 2-electron rate from SENSEI~\cite{Barak:2020fql}, while the shaded region includes bounds from XENON1T~\cite{Aprile:2019jmx}, LUX~\cite{Akerib:2018hck}, CRESST III~\cite{Abdelhameed:2019hmk} and CDEX~\cite{Liu:2019kzq}, as well as a recast of XENON10~\cite{Angle:2011th}, XENON100~\cite{Aprile:2016wwo}, and XENON1T~\cite{Aprile:2019xxb} data in terms of the Migdal effect~\cite{Essig:2019xkx}. For comparison, we show a projection for the Migdal effect in Xenon (dotted line) based on the atomic ionization signal~\cite{Essig:2019xkx}. In both panels, the shaded bands are an estimate of the theoretical uncertainty due to the impulse approximation, obtained by varying the threshold on $E_N$ from $4\bar\omega$ to $9\bar\omega$. See Appendix~\ref{app:rates} for more details. \label{fig:sensitivity}}
\end{figure*}

\mysection{Results\label{sec:results}}  The Migdal rate is given by $dR/d\omega = N_T n_\chi \int d^3 \bfv_\chi \, f(\bfv_\chi) \, v_\chi \, d\sigma/d\omega$, where $N_T$ is number of target nuclei per kilogram and $n_\chi = \rho_\chi/m_\chi$ is the DM number density. We take $\rho_\chi = 0.4 $ GeV/cm$^3$ and assume the standard halo model for the DM velocity distribution $f(\bfv)$ with escape velocity $v_{esc} = 500$ km/s \cite{Piffl:2013mla,Monari_2018} and Earth velocity $v_{e} = 240$ km/s.
We calculate the dielectric function with the public code \texttt{GPAW}~\cite{GPAW1, GPAW2, GPAW3, GPAW4}. The wavefunctions are computed on an $8\times8\times8$ $k$-space grid for Si, and a  $12\times12\times12$ grid for Ge. The TB09 exchange-correlation functional~\cite{TB09} is used and local field effects are incorporated~\cite{Adler}. A scissor correction is applied to match on to the experimentally-measured band gap in both materials. We have averaged the ELF over crystal directions, and for computational reasons, we currently only include the diagonal components of the loss function, although we verified that the off-diagonal terms do not contribute more than an $\mathcal{O}(1)$ amount to the total rate. Further details, more refined numerical studies and other applications will be presented in a forthcoming publication~\cite{futureKKL}. 

The resulting rates and sensitivity estimates are shown in Fig.~\ref{fig:sensitivity} for Si and Ge. The bands indicate an estimate of the theory uncertainty due the breakdown of the impulse approximation for low energy nuclear recoils. We find that our calculation starts to break down for $m_\chi\lesssim 50$ MeV, at which point one has to go beyond the impulse approximation by matching onto the phonon regime. The dashed lines in the right-hand panel indicate the free ion approximation, where the $F$ form factor is replaced with a $\delta$-function. We find that the free ion approximation is excellent in the regime where the impulse approximation applies, further validating the approach in \cite{Kozaczuk:2020uzb}. Compared to \cite{Kozaczuk:2020uzb}, we however find significantly stronger sensitivity by including the contributions away from the plasmon resonance. Most importantly, we confirm that the Migdal rate for a low threshold detector such as SENSEI, DAMIC or SuperCDMS is much higher than in noble liquid detectors. This is due to the lower ionization gap, the $\omega^{-4}$ scaling in \eqref{eq:Psemi_epsilon}, as well as the possibility of detecting all electronic excitations rather than only atomic ionizations.

\mysection{Note added} In the final stages of preparing this work, \cite{Kahn:2020fef} appeared, which studies the multiphonon response and has some overlap with the calculation in Appendix \ref{app:freeionapprox}.

\mysection{Acknowledgements}
We are grateful to Noah Kurinsky, Daniel Baxter, Gordan Krnjaic, Toby Opferkuch, and Chih-Pan Wu for useful discussions and to Yonatan Kahn for useful discussions and valuable comments on the draft. We further thank Diego Redigolo for useful discussions, proofreading the manuscript and collaboration on related work.
TL and JK are supported by the Department of Energy under grant DE-SC0019195 and a UC Hellman fellowship. TL is also supported by an Alfred P. Sloan foundation fellowship. 

\appendix

\section{Derivation of Migdal rate in semiconductors\label{app:semi_fullderivation}}

The derivation of the Migdal effect for semiconductors is complicated by the spatial delocalization of the valence electrons. As a consequence, each valence electron feels the presence of a large number of ions and their fellow electrons in the crystal. The system is often described in the single electron approximation, given by the Hamiltonian
\begin{equation}\label{eq:H0_crystal}
H_0 =  \frac{\bfp_e^2}{2 m_e} + U(\bfr)
\end{equation}
where $U(\bfr)$ is an effective, periodic potential felt by the electron, due to the presence of the ions and the remaining electrons.  In general $U(\bfr)$ is very complicated, and its eigenstates are typically obtained with specialized numerical methods in the realm of Density Functional Theory (DFT). For now, we can however keep $U(\bfr)$ as an abstract operator, and just work in terms of its eigenstates as long as possible. Concretely, the eigenstates of \eqref{eq:H0_crystal} are Bloch wave functions 
\begin{equation}\label{eq:bloch}
\psi_{j,\bfk }(\bfr)=\frac{1}{\sqrt{V}}u_{j,\bfk}(\bfr) \exp(i\bfk\cdot \bfr)
\end{equation}
where $j$ indexes the electronic bands. Going forward, we use $\ket{\bfk}$ and  $|\bfk]$ as shorthand for the full ($\psi_\bfk $) and Bloch functions ($u_{\bfk}$) respectively. To keep the notation manageable, the band indices $j$ will be often be suppressed in what follows. We quantize the system over a finite volume $V$ with periodic boundary conditions. The unit cell's volume is $\Omega$ and the number of cells in the crystal is therefore $N=V/\Omega$. We will take the infinite volume limit only at the end of the calculation. This means that the sampling over the first Brillouin zone (BZ) will be a finite, discrete sum with $N$ terms until we take the continuum limit by sending $V, N\to \infty$.

\begin{figure}[b]
\centering
\includegraphics[width=0.49\textwidth]{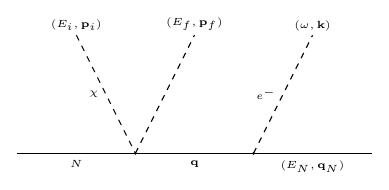}
\caption{Definition of kinematic variables.\label{fig:kinematics}}
\end{figure}

To treat the ion-electron interaction, we must account  for the screening from the spectator valence electrons, which is parameterized by $\epsilon$, the frequency-dependent, microscopic dielectric function. The dielectric function of a material is defined by the relation
\begin{equation}\label{eq:dielectricdef}
\bfE(\bfr,\omega)=\int \! d^3\bfr'\; \epsilon^{-1}(\bfr,\bfr',\omega)\bfE_{\text{ext}}(\bfr',\omega)
\end{equation}
with $\bfE$ and $\bfE_{\text{ext}}$ the total and external electric fields, respectively. For a homogeneous material at distances much larger than the lattice spacing, the dielectric function is invariant under translations, and its Fourier transform is a function of a single momentum vector, such that we can write it as $\epsilon(\bfq,\omega)$. At short distances, however, there can be large oscillations of the electric field inside the primitive cell, and translation invariance is broken down to the discrete translation vectors of the crystal. Using this residual translation invariance, one can show that the Fourier transform of \eqref{eq:dielectricdef} can be written as
\begin{equation}\label{eq:dielectricdeffourier}
\bfE(\bfk+\bfK,\omega)=\sum_{\bfK'} \epsilon^{-1}(\bfk+\bfK,\bfk+\bfK',\omega)\bfE_{\text{ext}}(\bfk+\bfK',\omega).
\end{equation}
where $\bfk$ is in the first BZ (1BZ), and $\bfK$ and $\bfK'$ are reciprocal lattice vectors. We will be using the common shorthand notation $\epsilon^{-1}_{\bfK\bfK'}(\bfk,\omega)\equiv\epsilon^{-1}(\bfk+\bfK,\bfk+\bfK',\omega)$. Note that $\epsilon^{-1}_{\bfK\bfK'}(\bfk,\omega)$ can be thought of as a matrix in the space of reciprocal lattice vectors, such that $\epsilon_{\bfK\bfK'}(\bfk,\omega)$ is defined as the matrix inverse of $\epsilon^{-1}_{\bfK\bfK'}(\bfk,\omega)$. Eventually we will drop terms for which $\bfK\neq\bfK'$, as they tend to be numerically subleading, and this distinction will not be important.

The Migdal rate is given by the leading order expansion in both the dark matter nucleus interaction ($H_\chi$) and the electromagnetic interaction ($H_e$), very analogously to a bremsstrahlung calculation. This is illustrated by the diagram in Fig.~\ref{fig:kinematics}, which also serves to define the notation for our kinematic variables. The corresponding interaction Hamiltonians are 
\begin{align}
 H_\chi&= \frac{2\pi b_\chi}{m_\chi} \delta(\bfr_\chi-\bfr_N)\\
 H_e&=- \int\!\! d^3\bfr' \frac{\Zion \alpha}{\epsilon(\bfr,\bfr',\omega)}\frac{1}{ | \bfr' - {\bf r}_{N}|}
 \label{eq:Hamquantum}
\end{align}
with $b_\chi$ the DM--nucleus scattering length and $\bfr_N, \bfr_\chi$ and $\bfr$ the position operators for the dark matter, nucleus and electronic wave functions respectively. Here we have assumed sub-GeV DM, so that the scattering length is related to the DM--nucleus elastic cross section by $\sigma_N=4\pi b_\chi^2 = A^2 \sigma_n$, where $\sigma_n$ is the DM--nucleon cross section and $A$ is mass number. $\alpha$ is the electromagnetic fine structure constant, and we have included the charges of the tightly bound core electrons in the charge of the ion, denoted by $\Zion$. In order to simplify the derivation below, we will simply keep a constant $\Zion$, and only restore the momentum-dependent $\Zion(k)$ in the final steps.

The Fourier transform of $H_e$ is 
\begin{align}
 H_e&=-  \frac{4\pi\alpha \Zion}{V}\sum_{\bfK,\bfK'} \sum_{\bfk}  \frac{\tilde\epsilon^{-1}_{\bfK\bfK'}(\bfk,\omega) e^{i(\bfk+\bfK)\cdot \bfr}}{|\bfk+\bfK||\bfk+\bfK'|}e^{-i(\bfk+\bfK')\cdot \bfr_N}\label{eq:Hamquantum}
\end{align}
where the sum over $\bfk$ samples the first Brillouin zone and $\bfK,\bfK'$ run over the reciprocal lattice vectors. We used the symmetrized dielectric matrix $ \tilde\epsilon^{-1}_{\bfK\bfK'}\equiv  \frac{|\bfk+\bfK|}{|\bfk+\bfK'|}\epsilon^{-1}_{\bfK\bfK'}$, which has the property that $\tilde\epsilon^{-1}_{\bfK\bfK'}=\tilde\epsilon^{-1}_{\bfK'\bfK}$. In what follows we will drop the $\tilde{\phantom{a}}$, and the $\epsilon_{\bfK\bfK'}^{-1}$ will always refer to the symmetrized dielectric matrix. 

There are two terms in the matrix element, corresponding to the advanced and retarded contributions, as in a standard bremsstrahlung calculation
\begin{align}
 \mathcal{M}_a&= \sum_{\bfq}  \frac{\bra{\bfq_N, \bfp_e+\bfk }H_e\ket{\bfq,\bfp_e}\bra {\bfp_f,\bfq } H_\chi \ket{\bfp_i,\psi_0}}{\omega+\frac{q_N^2}{2m_N} -\frac{q^2}{2m_N}}\label{eq:fullmatrixelementa}\\
  \mathcal{M}_b&=-\sum_{\bfq} \frac{\bra {\bfp_f,\bfq_N } H_\chi \ket{\bfq,\bfp_i }\bra{\bfq, \bfp_e+\bfk }H_e\ket{\bfp_e,\psi_0}}{\omega+\frac{q^2}{2m_N}}\label{eq:fullmatrixelementb}
 \end{align}
where the sum runs over the full momentum space for the ions.  As laid out in the main text and Appendix~\ref{app:freeionapprox}, our use of the impulse approximation means that we approximate both the final and the intermediate state nucleus wave function as plane waves, represented by $\ket{\bfq_N}$ and $\ket{\bfq}$ respectively.  The plane waves are normalized as $\bra{\bfr}\bfq\rangle=e^{i \bfq\cdot\bfr}/\sqrt{V}$,  etc. $\ket{\psi_0}$ indicates the ground state of the ion, which is in general not described as a plane wave, as it is bound in the crystal. We have neglected the small ground state energy in the propagator. We now calculate both terms separately.

First we calculate the DM--nucleus matrix element in \eqref{eq:fullmatrixelementa}, accounting for the fact that the nucleus is bound in the crystal.  Modeling both the incoming and outgoing the dark matter particle as a plane wave as well, the matrix element is
 \begin{align}
 \bra{\bfp_f,\bfq}  H_\chi \ket{\bfp_i, \psi_0 } &= \frac{2\pi b_\chi}{m_\chi}\frac{1}{V^{3/2}} \bra{ 0 } e^{i(\bfp_i-\bfp_f-\bfq)\cdot \bfr_N} \ket{\psi_0 }
 \end{align}
 with $\ket{0}$ the ground state of the free theory. The position space representation of the ground state wave function of a nucleus in a harmonic crystal is 
 \begin{equation}\label{eq:groundstate}
 \langle \bfr_N\ket{\psi_0}=\frac{( m_N^3 \bar\omega^3)^{1/4}}{\pi^{3/4}} e^{-\frac{1}{2}r_N^2 m_N \bar \omega}
 \end{equation}
with $\bar \omega$ the oscillator frequency, averaged with respect to the density of states $D(\omega)$ and the thermal Bose factor, or $\bar \omega\equiv \int_0^\infty\!\! d\omega' \,\omega' D(\omega')\coth(\frac{\omega'}{2T})$. We prove equation \eqref{eq:groundstate} in Appendix \ref{app:freeionapprox}. Evaluating the matrix element, one finds
\begin{align}
 \bra{\bfp_f,\bfq}  H_\chi \ket{\bfp_i, \psi_0  } = \frac{2\pi b_\chi}{m_\chi V^{3/2}}F(\bfp_i-\bfp_f-\bfq)\label{eq:nuclmatrixelement}
 \intertext{with}
 \label{eq:defF}
 F(\bfp_i-\bfp_f-\bfq)\equiv \left(\frac{4\pi}{m_N\bar\omega}\right)^{3/4}e^{\frac{-|\bfp_i-\bfp_f-\bfq|^2}{2m_N \bar\omega}}.
 \end{align}
In the limit of a free nucleus, we recover 
\begin{equation}
\lim_{\bar\omega\to 0} |F(\bfp_i-\bfp_f-\bfq)|^2=(2\pi)^3 \delta(\bfp_i-\bfp_f-\bfq)
\end{equation}
as expected. 

Next we calculate the electronic matrix element in \eqref{eq:fullmatrixelementa} by decomposing the electronic wave functions into the Bloch functions $|\bfp_e]$:
\begin{widetext}
\begin{align}
\bra{\bfq_N, \bfp_e+\bfk }H_e\ket{\bfq,\bfp_e}=& -\frac{4\pi\alpha \Zion}{V} \sum_{\bfK,\bfK'}\sum_{\bfk'} \epsilon^{-1}_{\bfK\bfK'}(\bfk',\omega) \frac{[\bfp_e+\bfk|e^{i(\bfk'+\bfK-\bfk)\cdot \bfr}|\bfp_e]}{|\bfk'+\bfK||\bfk'+\bfK'|}\bra{ \bfq_N } e^{-i(\bfk'+\bfK')\cdot\bfr_N}\ket{\bfq}\\
=& -\frac{4\pi \alpha\Zion}{V} \sum_{\bfK,\bfK'}\sum_{\bfk'} \epsilon^{-1}_{\bfK\bfK'}(\bfk',\omega)\frac{[\bfp_e+\bfk|e^{i(\bfk'+\bfK-\bfk)\cdot \bfr}|\bfp_e]}{|\bfk'+\bfK||\bfk'+\bfK'|}\delta_{\bfq, \bfq_N +\bfk'+\bfK'}\label{eq:elecmatrixelement}
\end{align}
where in the second line we used the plane wave forms to evaluate the matrix element corresponding to the nucleus. Note that we  use the Kronecker delta function for the ion wavefunction overlap here, and we will take $\delta_{\bfq, \bfq_N +\bfk'+\bfK'} \to ((2\pi)^3/V) \delta( \bfq_N +\bfk'+\bfK'-\bfq)$ in the continuum limit at the end. Next we can compute the Bloch amplitude, where we rewrite the amplitude as an integral just over the primitive cell, with volume $\Omega$:
\begin{align}
[\bfp_e+\bfk| e^{i(\bfk'+\bfK-\bfk)\cdot\bfr}  |\bfp_e]=& \frac{1}{V} \int_V d^3 \bfr\; u^\ast_{\bfp_e+\bfk}(\bfr) e^{i(\bfk'+\bfK-\bfk)\cdot\bfr} u_{\bfp_e}(\bfr)\\
=& \frac{1}{V} \int_\Omega d^3 \bfr\; u^\ast_{\bfp_e+\bfk}(\bfr) e^{i(\bfk'+\bfK-\bfk)\cdot\bfr} u_{\bfp_e}(\bfr)  \sum_\ell e^{i(\bfk'+\bfK-\bfk)\cdot\bfR_\ell}\\
=& \delta_{\bfk,\bfk'} \frac{1}{\Omega} \int_\Omega d^3 \bfr\; u^\ast_{\bfp+\bfk}(\bfr) e^{i\bfK\cdot\bfr} u_{ \bfp}(\bfr)\label{eq:blochmatrixelement}\\
\equiv& \delta_{\bfk,\bfk'}  [\bfp_e+\bfk| e^{i\bfK\cdot\bfr}| \bfp_e]_\Omega\label{eq:blochmatrixdef}
\end{align}
where the $\bfR_\ell$ are the lattice vectors of the Bravais lattice. The $\Omega$-subscript in \eqref{eq:blochmatrixdef} serves as a reminder that this matrix element is only defined over the volume of the primitive cell. In the second step we used the periodicity of the Bloch functions under translations along the Bravais lattice ($u_{j,\bfp}(\bfr+\bfR_\ell)=u_{j,\bfp}(\bfr)$). In \eqref{eq:blochmatrixelement} we furthermore relied on the identities $\sum_\ell e^{i\bfR_\ell\cdot(\bfk'-\bfk)}=\frac{V}{\Omega} \sum_{\bfL}\delta_{\bfk,\bfk'+\bfL}$ and $e^{i\bfR_\ell\cdot \bfK}=1$. Since both $\bfk$ and $\bfk'$ are restricted to the first Brillouin zone, the only possible contributions to the sum come from $\bfL=0$ and from the $\bfL$ in the second Brillouin zone. However, the latter configurations in $\bfk$ and $\bfk'$ are a set of measure zero in the $N \to \infty$ limit, so we drop them from the calculation. Inserting \eqref{eq:blochmatrixdef} back into \eqref{eq:elecmatrixelement} results in
\begin{align}
&\bra{\bfq_N, \bfp_e+\bfk }H_e\ket{\bfq,\bfp_e}= -\frac{4\pi \alpha\Zion}{V} \sum_{\bfK,\bfK'} \epsilon^{-1}_{\bfK\bfK'}(\bfk,\omega)\frac{[\bfp_e+\bfk| e^{i\bfr\cdot\bfK}| \bfp_e]_\Omega}{|\bfk+\bfK||\bfk+\bfK'|}\delta_{\bfq, \bfq_N +\bfk+\bfK'}\label{eq:elecmatrixelement2}
\end{align}
and thus
\begin{align}
 \mathcal{M}_a&= -\frac{4\pi \alpha\Zion}{V^{5/2}}\frac{2\pi b_\chi}{m_\chi} \sum_{\bfK,\bfK'} \epsilon^{-1}_{\bfK\bfK'}(\bfk,\omega)\frac{[\bfp_e+\bfk| e^{i\bfr\cdot\bfK}| \bfp_e]_\Omega}{|\bfk+\bfK||\bfk+\bfK'|}  \frac{F(\bfp_i-\bfp_f-\bfq_N -\bfk-\bfK')}{\omega -\frac{\bfq_N\cdot(\bfk+\bfK')}{m_N}-\frac{|\bfk+\bfK'|^2}{2m_N}}\label{eq:Mafinal}
 \end{align}

The calculation of $\mathcal{M}_b$ is analogous: The $H_\chi$ element in \eqref{eq:fullmatrixelementb} is trivial, as all incoming and outgoing states are plane waves
\begin{equation}
\bra {\bfp_f,\bfq_N } H_\chi \ket{\bfq,\bfp_i }=\frac{2\pi b_\chi}{m_\chi}\frac{1}{V}\delta_{\bfp_f+\bfq_N,\bfp_i+\bfq}.
\end{equation}
Using the same methods as above, the $H_e$ matrix element is 
\begin{equation}
\bra{\bfq, \bfp_e+\bfk }H_e\ket{\bfp_e,\psi_0}= -\frac{4\pi \alpha\Zion}{V^{3/2}} \sum_{\bfK,\bfK'} \epsilon^{-1}_{\bfK\bfK'}(\bfk,\omega)\frac{[\bfp_e+\bfk| e^{i\bfr\cdot\bfK}| \bfp_e]_\Omega}{|\bfk+\bfK||\bfk+\bfK'|}F(\bfq+\bfk+\bfK')
\end{equation}
such that
\begin{align}
  \mathcal{M}_b&=\frac{4\pi\alpha\Zion}{V^{5/2}} \frac{2\pi b_\chi}{m_\chi}\sum_{\bfK,\bfK'}  \epsilon^{-1}_{\bfK\bfK'}(\bfk,\omega)\frac{[\bfp_e+\bfk| e^{i\bfr\cdot\bfK}| \bfp_e]_\Omega}{|\bfk+\bfK||\bfk+\bfK'|}
 \frac{F(\bfp_i-\bfp_f-\bfq_N -\bfk-\bfK')}{\omega+\frac{|\bfk+\bfK'|^2}{2m_N}}.\label{eq:Mbfinal}
\end{align}

Adding both contributions in \eqref{eq:Mafinal} and \eqref{eq:Mbfinal}, the total matrix element is 
\begin{align}
  \mathcal{M}&=-\frac{4\pi \alpha\Zion}{V^{5/2}} \frac{2\pi b_\chi}{m_\chi}\sum_{\bfK,\bfK'}  \epsilon^{-1}_{\bfK\bfK'}(\bfk,\omega)\frac{[\bfp_e+\bfk| e^{i\bfr\cdot\bfK}| \bfp_e]_\Omega}{|\bfk+\bfK||\bfk+\bfK'|}F(\bfp_i-\bfp_f-\bfq_N -\bfk-\bfK')\left[\frac{1}{\omega -\frac{\bfq_N\cdot(\bfk+\bfK')}{m_N}}-\frac{1}{\omega}\right]\label{eq:Mfinal}
\end{align}
where we have dropped the negligible $|\bfk+\bfK'|^2/2m_N$ terms from the propagators. This is justified, as the form factor $F$ exponentially suppresses contributions from terms with $\bfK' \gg \bfp_i-\bfp_f-\bfq_N$.

The cross section is obtained as usual by invoking Fermi's golden rule
\begin{align}
\sigma =& \frac{V}{v_\chi} 2\pi \int\!\!d\omega\, \sum_{j,\bfp_e}\sum_{\substack{\bfq_N,\bfp_f\\j',\bfk}} | \mathcal{M}|^2 f_j(\bfp_e)\big(1-f_{j'}(\bfp_e+\bfk)\big)\delta\left({E_i-E_f-\omega-\frac{q^2_N}{2m_N}}\right)\delta(\omega_{j',\bfp_e+\bfk}-\omega_{j,\bfp_e}-\omega)\\
=& \frac{V}{v_\chi} 2\pi \int\!\!d\omega\, \sum_{j,\bfp_e}\sum_{\substack{\bfq_N,\bfp_f\\j',\bfk}} | \mathcal{M}|^2 \frac{f_j(\bfp_e)-f_{j'}(\bfp_e+\bfk)}{1-e^{-\omega/T}}\delta\left({E_i-E_f-\omega-\frac{q^2_N}{2m_N}}\right)\delta(\omega_{j',\bfp_e+\bfk}-\omega_{j,\bfp_e}-\omega)\label{eq:detailedbalance}
 \end{align} 
with $v_\chi$ is the incoming DM velocity and $f_j(\bfp_e)$ the temperature-dependent occupation numbers for the electronic bands, which are indexed by $j,j'$. We have also introduced an integral over $\omega$ with a corresponding $\delta$-function, such that we can continue to use the shorthand notation $\omega=\omega_{j',\bfp_e+\bfk}-\omega_{j,\bfp_e}$. In the second line we rewrote the Pauli blocking factor by identifying the $f_j(\bfp_e)$ as the Fermi-Dirac distribution with temperature $T$. CCD detectors such as DAMIC and SENSEI are run at $T\approx 140$ K, while SuperCDMS is operated at even lower temperatures. This means that for $\omega \sim $ eV we can safely neglect the $e^{-\omega/T}$ factor in \eqref{eq:detailedbalance}. 

Next we take the infinite volume limit, which in momentum space corresponds to the continuum limit, by replacing $\sum_{\bfk}\to V \int \!\!\frac{d^3\bfk}{(2\pi)^3}$, etc. Plugging in \eqref{eq:Mfinal}, the differential cross section is thus
 \begin{align}
\frac{d\sigma}{d\omega} =& \frac{8\pi \alpha \Zion^2}{v_\chi} \left(\frac{2\pi b_\chi}{m_\chi}\right)^2  \int\!\frac{d^3\bfq_N}{(2\pi)^3}\int\!\frac{d^3\bfp_f}{(2\pi)^3}\int\!\!\frac{d^3\bfk}{(2\pi)^3}\sum_{\bfK,\bfK'}\sum_{\bfL,\bfL'}  \epsilon^{-1}_{\bfK\bfK'}(\bfk,\omega)\epsilon^{-1\ast}_{\bfL\bfL'}(\bfk,\omega)\nonumber\\
&\times\frac{1}{|\bfk+\bfK'||\bfk+\bfL'|} \left[\frac{1}{\omega -\frac{\bfq_N\cdot(\bfk+\bfK')}{m_N}}-\frac{1}{\omega}\right]\left[\frac{1}{\omega -\frac{\bfq_N\cdot(\bfk+\bfL')}{m_N}}-\frac{1}{\omega}\right]\nonumber\\
& \times F(\bfp_i-\bfp_f-\bfq_N -\bfk-\bfK') F(\bfp_i-\bfp_f-\bfq_N -\bfk-\bfL')\,\delta\left({E_i-E_f-\omega-\frac{q^2_N}{2m_N}}\right) \nonumber \\
&\times\frac{4\pi^2\alpha}{V}\sum_{j,j',\bfp_e} \frac{[\bfp_e+\bfk| e^{i\bfr\cdot\bfK}| \bfp_e]_\Omega[\bfp_e| e^{-i\bfr\cdot\bfL}| \bfp_e+\bfk]_\Omega}{|\bfk+\bfK||\bfk+\bfL|} (f_j(\bfp_e)-f_{j'}(\bfp_e+\bfk)) \, \delta(\omega_{j',\bfp_e+\bfk}-\omega_{j,\bfp_e}-\omega) 
\label{eq:rateintermediate}
\end{align}
where the last line in \eqref{eq:rateintermediate} equals $\text{Im}[\epsilon_{\bfK\bfL}]$ in the random phase approximation (RPA). This follows from the generalization of Lindhard's formula \cite{osti_4405425} to the reciprocal lattice \cite{PhysRev.126.413}\footnote{The formula differs from Ref.~\cite{PhysRev.126.413} with a $|\bfk+\bfK|/|\bfk+\bfL|$ factor, as we are working with the symmetrized dielectric function.}
\begin{equation}
\epsilon_{\bfK\bfL}(\bfq,\omega)=\delta_{\bfK,\bfL} - \frac{4\pi\alpha}{V}\lim_{\eta\to0^+}\sum_{j,j',\bfp_e} \frac{[\bfp_e+\bfk| e^{i\bfr\cdot\bfK}| \bfp_e]_\Omega[\bfp_e| e^{-i\bfr\cdot\bfL}| \bfp_e+\bfk]_\Omega}{|\bfk+\bfK||\bfk+\bfL|} \frac{f_j(\bfp_e)-f_{j'}(\bfp_e+\bfk)}{\omega_{j',\bfp_e+\bfk}-\omega_{j,\bfp_e}-\omega-i\eta} .
\end{equation}
Note that the transition matrix elements are real for systems respecting time reversal and parity symmetry \cite{Deslippe_2012}, such that the only contribution to $\text{Im}[\epsilon_{\bfK\bfL}]$ is from the pole in the propagator.
One can moreover show that $\sum_{\bfK,\bfL}\epsilon^{-1}_{\bfK\bfK'}(\bfk,\omega)\text{Im}[\epsilon_{\bfK\bfL}(\bfk,\omega)]\epsilon^{-1\ast}_{\bfL\bfL'}(\bfk,\omega)=\text{Im}[- \epsilon^{-1}_{\bfK'\bfL'}(\bfk,\omega)]$, such that we find
\begin{align}
\frac{d\sigma}{d\omega} =& \frac{8\pi \Zion^2 \alpha }{v_\chi} \left(\frac{2\pi b_\chi}{m_\chi}\right)^2 \, \int\!\frac{d^3\bfq_N}{(2\pi)^3}\int\!\frac{d^3\bfp_f}{(2\pi)^3}\int\!\!\frac{d^3\bfk}{(2\pi)^3}\sum_{\bfK,\bfL}
\frac{\text{Im}[- \epsilon^{-1}_{\bfK\bfL}(\bfk,\omega)]}{|\bfk+\bfK||\bfk+\bfL|} \left[\frac{1}{\omega -\frac{\bfq_N\cdot(\bfk+\bfK)}{m_N}}-\frac{1}{\omega}\right]\left[\frac{1}{\omega -\frac{\bfq_N\cdot(\bfk+\bfL)}{m_N}}-\frac{1}{\omega}\right]\nonumber\\
& \times F(\bfp_i-\bfp_f-\bfq_N -\bfk-\bfK) F(\bfp_i-\bfp_f-\bfq_N -\bfk-\bfL)\,\delta\left({E_i-E_f-\omega-\frac{q^2_N}{2m_N}}\right)
\intertext{The $\text{Im}[- \epsilon^{-1}_{\bfK\bfL}]$ object is a positive function and corresponds to the well known energy loss function (ELF), which governs the energy loss of charged particles passing through a material. To reduce the computational burden, we currently set $\text{Im}[- \epsilon^{-1}_{\bfK\bfL}]=0$ for $\bfK\neq\bfL$, although we verified that the combined, neglected contributions are at most comparable to the diagonal terms. With this additional assumption, we arrive at }
\frac{d\sigma}{d\omega}=&\frac{8\pi \alpha }{v_\chi} \left(\frac{2\pi b_\chi}{m_\chi}\right)^2 \int\!\frac{d^3\bfq_N}{(2\pi)^3}\int\!\frac{d^3\bfp_f}{(2\pi)^3}\int\!\!\frac{d^3\bfk}{(2\pi)^3}\sum_{\bfK} \Zion^2(| \bfk + \bfK |) \, \frac{\text{Im}[- \epsilon^{-1}_{\bfK\bfK}(\bfk,\omega)]}{|\bfk+\bfK|^2}   \left[\frac{1}{\omega -\frac{\bfq_N\cdot(\bfk+\bfK)}{m_N}}-\frac{1}{\omega}\right]^2 \nonumber\\
&\times |F(\bfp_i-\bfp_f-\bfq_N -\bfk-\bfK)|^2\, \delta\left({E_i-E_f-\omega-\frac{q^2_N}{2m_N}}\right). \label{eq:ratefinal}
 \end{align}
Note that in this expression, we have finally have restored the effective ion charge $\Zion(| \bfk + \bfK |)$ associated with the momentum transfer from the ion to the electrons, $\bfk + \bfK$.

In \eqref{eq:ratefinal}, all information regarding the material and its properties is encoded in the reciprocal lattice vectors, the ELF and the form factor $F(\bfq)$, which is sensitive to the phonon density of states (see \eqref{eq:defF}). Formulating the rate for the Migdal process in terms of the ELF and the phonon density of states has a number of advantages. Both can be measured directly for individual materials under various experimental conditions. A large body of experimental data is already available, though primarily for $\bfk\approx 0$ in the case of the ELF. It is therefore often necessary to extrapolate the experimental data on the ELF to finite $\bfk$, which is can be done with Lindhard or Mermin oscillator models \cite{PhysRevB.1.2362,VOS2019242}. In fact, Ref.~\cite{Liu:2020pat} already proposed to use optical data to estimate the rate for the Migdal effect in various materials. Our calculations put this approach on a more firm theoretical footing. In particular, \cite{Liu:2020pat} is relying on \eqref{eq:M_boost}, which we have shown to be invalid for semiconductors. In addition, both the ELF and the density of states can be directly computed from first principles with well established, public codes such as \texttt{GPAW} and \texttt{Quantum ESPRESSO}. This has the advantage that no extrapolation to finite $\bfk$ is needed, though this method is substantially more computationally intensive. We explore both techniques further in a companion paper~\cite{futureKKL}, but illustrate the advantage of this formalism in Fig.~\ref{fig:Si_ELF} which compares the ELF in Si computed in the DFT code \texttt{GPAW} to that measured from data. The ELF for $\bfk'$ outside the 1BZ is defined as Im$[-\epsilon^{-1}_{\bfK \bfK}(\bfk,\omega)]$ where $\bfk'=\bfk + \bfK$ with $\bfk \in$ 1BZ and $\bfK$ a reciprocal lattice vector. The results shown are for momentum transfer along the [111] direction. The DFT calculation reproduces the data well, and comparison with data allows for more control over theoretical uncertainties in the Migdal rate calculation.

\begin{figure*}[t]
\includegraphics[width=\textwidth]{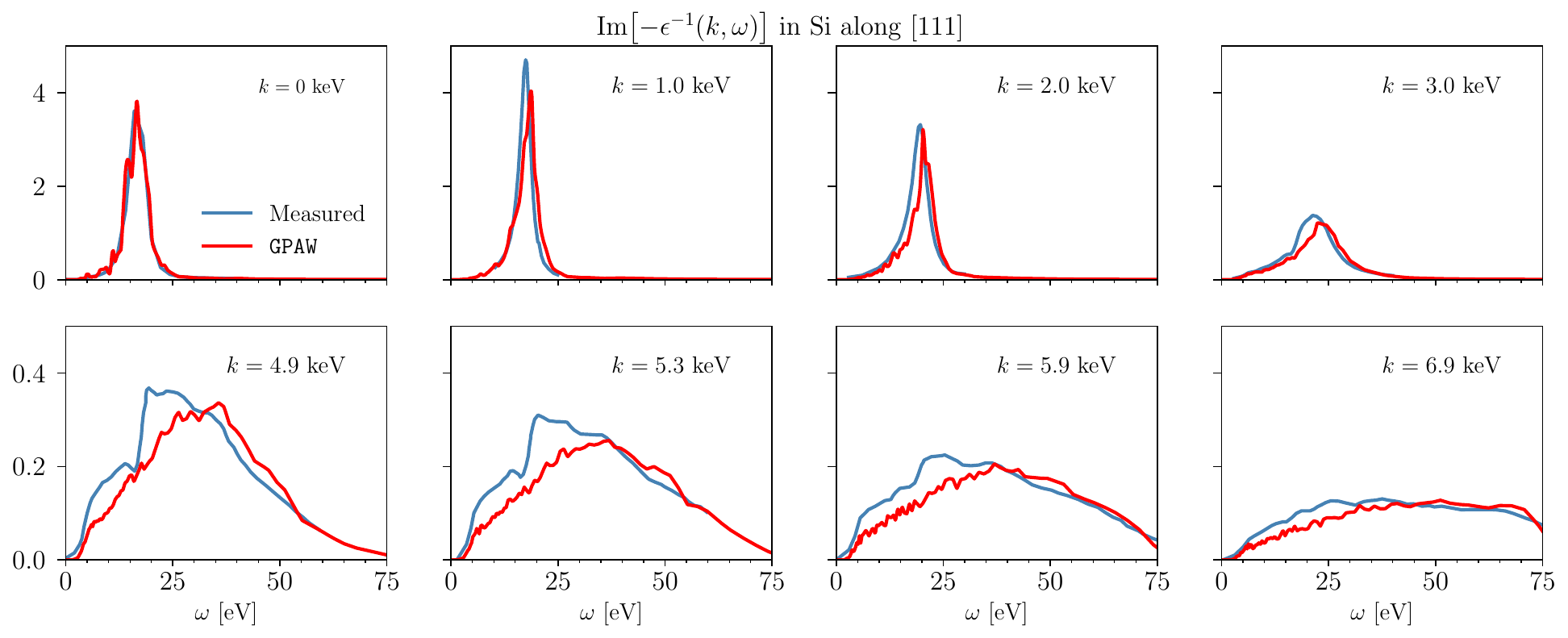}
\caption{Energy loss function in Silicon for various $k$ values. The red curves show the result computed in \texttt{GPAW} using the TB09 exchange correlation functional with scissor-corrected bandgap, while the blue curves show the corresponding measurement from optical data ($k=0$)~\cite{Optical} and  inelastic x-ray scattering measurements ($k>0$)~\cite{Weissker}. The DFT calculation reproduces the experimental results well. For low $k$, the plasmon peak is clearly visible, while it disappears for high $k$. \label{fig:Si_ELF}}
\end{figure*}
It is possible to further simplify the expression above by making a number of additional approximations, which provide some conceptual insight and simplify the numerical evaluation of the integrals in \eqref{eq:ratefinal}. First, we observe that in most of the phase space $\bfq_N\cdot(\bfk+\bfK)/m_N\ll \omega$ and $|\bfk+\bfK|\ll|\bfq_N|$, which corresponds to the so-called soft limit. This allows us to factorize the various integrals as follows
\begin{align}
\frac{d\sigma}{d\omega}\approx&\frac{8\pi \alpha }{v_\chi} \left(\frac{2\pi b_\chi}{m_\chi}\right)^2 \int\!\frac{d^3\bfq_N}{(2\pi)^3}\int\!\frac{d^3\bfp_f}{(2\pi)^3} |F(\bfp_i-\bfp_f-\bfq_N)|^2\, \delta\left({E_i-E_f-\omega-\frac{q^2_N}{2m_N}}\right)\nonumber\\
&\times\int\!\!\frac{d^3\bfk}{(2\pi)^3}\sum_{\bfK} \Zion^2(| \bfk + \bfK|) \, \frac{\text{Im}[- \epsilon^{-1}_{\bfK\bfK}(\bfk,\omega)]}{|\bfk+\bfK|^2}   \frac{|\bfq_N\cdot(\bfk+\bfK)|^2}{\omega^4m_N^2} \label{eq:ratefinalsoft}
\end{align}
This is the formula we use to obtain the numerical results shown in Fig.~\ref{fig:sensitivity}. The free ion approximation is obtained by replacing the $|F|^2$ form factor with a $\delta$-function and subsequently eliminating the $\bfp_f$ integral
\begin{align}
\frac{d\sigma}{d\omega}\approx&\frac{8\pi \alpha }{v_\chi} \left(\frac{2\pi b_\chi}{m_\chi}\right)^2 \int\!\frac{d^3\bfq_N}{(2\pi)^3}\, \delta\left({E_i-E_f-\omega-\frac{q^2_N}{2m_N}}\right)\int\!\!\frac{d^3\bfk}{(2\pi)^3}\sum_{\bfK} \Zion^2(| \bfk + \bfK|) \, \frac{\text{Im}[- \epsilon^{-1}_{\bfK\bfK}(\bfk,\omega)]}{|\bfk+\bfK|^2}   \frac{|\bfq_N\cdot(\bfk+\bfK)|^2}{\omega^4m_N^2} \label{eq:freeionfinal}
\end{align}
 To make contact with earlier analytic work, we can further restrict the phase space to $\bfK=0$, such that this expression reduces to 
\begin{align}
\frac{d\sigma}{d\omega} \approx&\frac{8\pi \alpha }{v_\chi}\left(\frac{2\pi b_\chi}{m_\chi}\right)^2 \int\!\frac{d^3\bfq_N}{(2\pi)^3}\delta(E_i-E_f-\omega-\frac{q^2_N}{2m_N})\int\!\!\frac{d^3\bfk}{(2\pi)^3} \frac{\Zion^2(k)}{k^2}\text{Im}\left[-\epsilon^{-1}_{00}(\bfk,\omega)\right] \frac{|\bfv_N\cdot \bfk|^2}{\omega^4} .\label{eq:ratefinalsemiclassical}
\end{align}
which matches the semi-classical result derived in \cite{Kozaczuk:2020uzb} with fixed $\Zion$. Now recall that $\frac{d\sigma}{d\omega E_N}\approx \frac{d\sigma_{\text{el}}}{dE_N}\times\frac{dP}{d\omega}$ in the soft limit, with
\begin{align}
    \frac{d\sigma_{\text{el}}}{dE_N} = \frac{A^2 \sigma_n m_N }{2 \mu_{\chi n}^2 v_\chi^2} \times \Theta\left(v_\chi -  \frac{1}{\sqrt{2 m_N E_N}} \left( \frac{m_N E_N}{\mu_{N\chi}} + \omega \right) \right),
\end{align}
and for sub-GeV dark matter we can take $\mu_{\chi n}, \mu_{\chi N} \to m_\chi$.
After carrying out the $\bfq_N$ integral in \eqref{eq:ratefinalsemiclassical} we can read off the electronic excitation probability 
\begin{align}
\frac{dP}{d\omega}=&4\alpha\int\!\!\frac{d^3\bfk}{(2\pi)^3} \frac{\Zion^2(k)}{k^2}\text{Im}\left[-\epsilon^{-1}_{00}(\bfk,\omega)\right]\frac{|\bfv_N\cdot \bfk|^2}{\omega^4}.
\end{align}
To make the connection with the Migdal effect in atoms in \eqref{eq:atomic_fourier} as manifest as possible, we can substitute the matrix element back for $\text{Im}[\epsilon(\bfk,\omega)]$ 
\begin{align}
\frac{dP}{d\omega}=&4\alpha^2  \frac{4\pi^2}{V}\sum_{\bfp_e}\int\!\!\frac{d^3\bfk}{(2\pi)^3} \frac{\Zion^2(k)}{k^4}\frac{|\bfv_N\cdot \bfk|^2}{\omega^4} \frac{\left|[\bfp_e+\bfk |\bfp_e]_\Omega \right|^2}{|\epsilon(\bfk,\omega)|^2 } \left(f(\bfp_e)-f(\bfp_e+\bfk)\right)\delta(\omega_{\bfp_e+\bfk}-\omega_{\bfp_e} -\omega)
\end{align}
where $\epsilon(\bfk,\omega)$ is the macroscopic dielectric function.
\end{widetext}

\section{Multiphonon response and impulse approximation\label{app:freeionapprox}}

In this Appendix we compute the response associated with dark matter scattering off a nucleus which is embedded in an isotropic, harmonic crystal, to all orders in the phonon expansion. This is the generalization of the elastic nuclear recoil calculation, without the Migdal effect. Analogous to the analysis in \cite{gunnwarner}, our calculation allows us to interpolate between the single phonon regime at low momentum transfers, and the free ion regime at high momentum transfers. Our goals hereby are two-fold: \emph{(i)} We explicitly demonstrate that the impulse approximation is an excellent approximation to the full multiphonon response when the dimensionless parameter $q/\sqrt{2m_N\bar\omega}\gtrsim 1$, where $q$ is the momentum imparted on the nucleus and $\bar\omega$ the typical frequency of the phonons. \emph{(ii)} We show that the response in this regime is well modeled by using the ground state of the harmonic system for the initial state of the nucleus, along with the plane wave approximation for its final state, justifying the approach in Appendix~\ref{app:semi_fullderivation}.

For a short range interaction, the interaction between the nucleus and the DM is simply
\begin{equation}\label{eq:neutronpotential}
\mathcal{V}(\bfr )= V_0 \delta(\bfr _N-\bfr ) \rightarrow   \tilde V(\bfq) = \tilde V_0 e^{i \bfq \cdot \bfr_N}.
\end{equation}
with $\tilde V$ the Fourier transform of the potential. Here we are only concerned with establishing the momentum transfer for which the free ion approximation breaks down, rather than the overall normalization. We will therefore set the constant prefactor $\tilde V_0=  1$ going forward. The dynamical structure factor describes the response of the lattice, and can be defined by Fermi's Golden rule:
\begin{align}\label{skodef}
S(\bfq,\omega)&\equiv  \sum_{\lambda_f} \left|\bra {\lambda_f}e^{-i \bfq\cdot \bfr_N} \ket{\lambda_i}\right|^2 \delta (E_{\lambda_f}-E_{\lambda_i} -\omega)
\intertext{with $\bra{\lambda_{i,f}}$ the initial and final states respectively. The expectation value includes a thermal average over initial states. Upon Fourier transforming the energy $\delta$-function, this can be written as}
&=\frac{1}{2\pi}\int_{-\infty}^{+\infty}\!\! dt\,\sum_{\lambda_f}\bra {\lambda_i } e^{-i\bfq\cdot\bfr _{N}} \ket{\lambda_f}\nonumber\\
&\qquad\times\bra{\lambda_f } e^{i E_{\lambda_f}t}e^{i\bfq\cdot\bfr_N} e^{-i E_{\lambda_i}t}\ket{\lambda_i} e^{-i\omega t}\\
\intertext{Given that the $\lambda_{i,f}$ form a complete basis of eigenstates of the Hamiltonian, this can be expressed in terms of a time-dependent two-point correlation function}
&=\frac{1}{2\pi}\int_{-\infty}^{+\infty}\!\! dt\,\langle e^{-i\bfq\cdot\bfr _{N}(0)}  e^{i\bfq\cdot\bfr_N(t)} \rangle e^{-i\omega t}
\end{align}
where in the last line we dropped the explicit reference to the initial states $\bra{\lambda_i}$.  With Bloch's identity and a moderate amount of matrix algebra, the expectation value can be taken into the exponent (see e.g.~Appendix B of \cite{Griffin:2018bjn} or chapter 9 of \cite{Schober2014})
\begin{align}\label{skofinal}
S(\bfq,\omega)=\frac{1}{2\pi}\int_{-\infty}^{+\infty}\!\!\!\!\!\!\! dt\, e^{-2 W(\bfq)} e^{\langle\bfq\cdot\bfr _{N}(0)\,\bfq\cdot\bfr_N(t)\rangle}  e^{-i\omega t}
\end{align}
with $W(\bfq)\equiv\frac{1}{2}\langle(\bfq\cdot\bfr _{N}(0))^2\rangle$ the Debye-Waller factor. 

In the small $\bfq$ limit, one can expand the time-dependent exponential to extract the DM scattering rate to a single phonon \cite{Griffin:2018bjn,Trickle:2019nya} and two phonons \cite{Campbell-Deem:2019hdx}. Here we are however interested in the full momentum range, and cannot expand the exponential. Instead, we first compute the correlation function exactly for the special case of an isotropic harmonic oscillator with frequency $\omega_0$. In particular, the second quantization of the displacement operator is given by
\begin{equation}\label{eq:secondquant}
\bfr_N(t)=\frac{1}{\sqrt{2m_N \omega_0}}\sum_j \bfe_j \left( a\, e^{i \omega_0 t} +a^\dagger e^{-i \omega_0 t}\right)
\end{equation}
where the $\bfe_j$ are a set of three unit vectors in position space and $a$ ($a^\dagger$) are creation (annihilation) operators.  All directions are weighted equally since we assumed the isotropic limit. Here we also implicitly assumed that the equilibrium location of the ion is at the origin of the coordinate frame; more general expressions can be found in Appendix B of \cite{Griffin:2018bjn} or in \cite{Schober2014}. With \eqref{eq:secondquant} in hand, we can calculate
\begin{align}
\langle\bfq\cdot\bfr _{N}(0)\,\bfq\cdot\bfr_N(t)\rangle &= \frac{q^2}{2 m_N \omega_0}\Big[[(1+n(\omega_0)
)e^{i \omega_0 t}\nonumber\\
&\quad\quad+n(\omega_0)e^{-i \omega_0 t}\Big]
\end{align}
with $n(\omega_0)\equiv1/(e^{\omega_0/T}-1)$ the Bose-Einstein distribution for a temperature $T$. The first and second term respectively correspond to the absorption and production of an oscillatory mode. This can be further rewritten as
\begin{align}\label{eq:corrHarmosc}
\langle\bfq\cdot\bfr _{N}(0)\,\bfq\cdot\bfr_N(t)\rangle &= \frac{q^2}{2 m_N \omega_0}\Big[\cos(\omega_0 t)\coth\left(\frac{\omega_0}{2 T}\right)\nonumber\\
 &\quad\quad+i \sin(\omega_0 t)\Big]
 \end{align}
 The Debye-Waller factor is obtained by setting $t=0$
 \begin{equation}
W(q)= \frac{1}{2}\frac{q^2}{2 m_N \omega_0}\coth\left(\frac{\omega_0}{2 T}\right)
\end{equation}

In a realistic crystal, there is of course a continuum of frequencies, rather than a single frequency $\omega_0$.\footnote{In the special case of the harmonic oscillator, the energy spectrum is discrete instead of continuous. At high momentum transfer, the spectral lines however blur together if the energy resolution is finite, and the gaussian form one obtains in the impulse approximation is to be understood as the envelope of the narrowly spaced spectral lines. See e.g.~\cite{watson1996neutron} for details.} This can be accounted for by including a density of states $D(\omega)$, such that \eqref{eq:corrHarmosc} generalizes to 
\begin{align}\label{eq:corrCrystal}
\langle\bfq\cdot&\bfr _{N}(0)\,\bfq\cdot\bfr_N(t)\rangle = \frac{q^2}{2 m_N}\int\!\! d\omega' \frac{D(\omega')}{\omega'}\nonumber\\
 &\times \Big[\cos(\omega' t)\coth\left(\frac{\omega'}{2T}\right)+i \sin(\omega' t)\Big]
\end{align}
and similarly for the Debye-Waller function.
While this system is in general no longer integrable, the frequency and time integrals in \eqref{skofinal} and \eqref{eq:corrCrystal} can be carried out numerically for a given $D(\omega)$. To illustrate the qualitative features of the transition to the free ion regime, we work with a simple, isotropic Debye density of states defined by
\begin{equation}\label{eq:debeye}
D(\omega)=\left\{\begin{array}{ll} 3|\omega|^2/\omega_{ph}^3&|\omega|\leq\omega_{ph}\\ 0&|\omega|>\omega_{ph}\end{array}\right.
\end{equation}
where $\omega_{ph}$ is the frequency of the acoustic phonons near the edge of the first Brillouin zone.\footnote{Optical phonons tend to have higher frequencies, however their coupling to the dark matter is strongly suppressed in mono-atomic materials such as Si, Ge and diamond \cite{Knapen:2017ekk,Griffin:2018bjn,Cox:2019cod}.} For Si and Ge, $\omega_{ph}$ can taken to be roughly 40 meV and 25 meV respectively.

In the high $\bfq$ regime, the response will be dominated by short time scales. An asymptotic expression can be derived through a steepest-descent expansion of \eqref{skofinal} around small $t$ \cite{gunnwarner}. This corresponds to the impulse approximation, as the superscripts indicate
\begin{equation}\label{eq:sko_impulseapprox}
S^{IA}(\bfq,\omega) = \frac{1}{\sqrt{2\pi\Delta^2}}e^{-\frac{\left(\omega - \frac{q^2}{2m_N}\right)^2}{2\Delta^2}}
\end{equation}
 with $ \Delta^2\equiv\frac{q^2\bar \omega}{2m_N}$ and 
 \begin{align}
\bar \omega&\equiv \int_{-\infty}^{+\infty}\!\!\!\! d\omega' \,\omega' D(\omega')\frac{1}{e^{\omega'/T}-1}\\
&=\int_0^{+\infty}\!\!\!\! d\omega' \,\omega' D(\omega')\coth\left(\frac{\omega'}{2T}\right)\label{eq:averagedomega}
\end{align}
where $\bar\omega$ is the phonon frequency, averaged over the density of states and the thermal Bose-factor. (Negative frequencies correspond to the dark matter absorbing energy from the crystal.) When $T=0$, $\bar\omega = 3\omega_{ph}/4$ for the density of states in \eqref{eq:debeye}. In the free limit of $\bar\omega\to0$, or equivalently $q\to \infty$, \eqref{eq:sko_impulseapprox} reduces to the familiar $\delta$-function, enforcing the on-shell condition of an unbound nucleus.

\begin{figure*}[t]
\includegraphics[width=\textwidth]{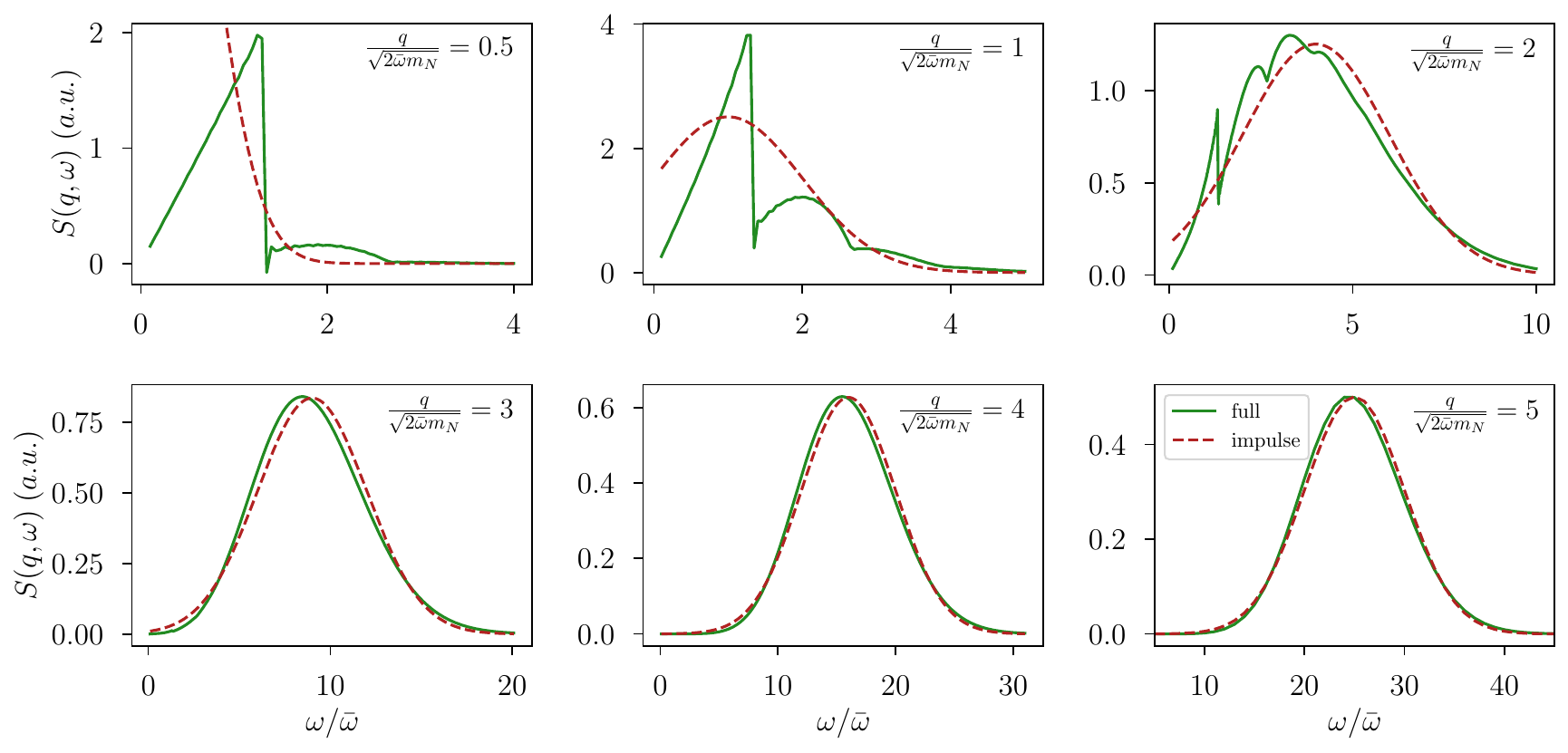}
\caption{The dynamic structure factor $S(q,\omega)$ encodes the response of the crystal to nuclear interactions. Here we show $S(q,\omega)$ in arbitrary units for different values of the momentum transfer, assuming an isotropic, harmonic crystal with a Debye density of states and $T=0$. The full result and impulse approximation are shown respectively with the solid and dashed curves. At large $q/\sqrt{2\bar\omega m_N}$, the structure factor resembles a Gaussian which is centered at the free-particle recoil energy $q^2/(2m_N)$ and with width $\sqrt{\bar \omega q^2/(2m_N)}$.  \label{fig:freeion}}
\end{figure*}

In Fig.~\ref{fig:freeion}, our numerical evaluation of \eqref{skofinal} is compared with the impulse approximation in \eqref{eq:sko_impulseapprox}, for various values of the momentum transfer. For $q/\sqrt{2m_N\bar \omega}=0.5$, the rate is dominated by the production of a single phonon, and the structure factor simply reflects the phonon density of states $D(\omega)$. Though the impulse approximation is clearly not valid, expanding the exponential factor in \eqref{skofinal} in low $q$ is however an excellent approximation. This regime was studied in \cite{Knapen:2017ekk,Griffin:2018bjn,Griffin:2019mvc,Trickle:2019nya,Cox:2019cod,Campbell-Deem:2019hdx,Griffin:2020lgd}. For $q/\sqrt{2m_N\bar \omega}\approx 1$, a bump at higher frequencies starts to emerge, which corresponds to two phonon production. For even higher $q$, the peaks associated with single, double and triple phonon production are clearly visible, as well as a hint of the four phonon bump. Finally at $q/\sqrt{2m_N\bar \omega}\gtrsim 3$, the individual phonon peaks are no longer discernible, and the structure function has converged to the Gaussian form in \eqref{eq:sko_impulseapprox}, associated with the impulse approximation. The slight offset of the maximum between both results is due to the fact that at high $q$, the true saddle point in \eqref{skofinal} is slightly offset from $t=0$, an effect which can be corrected for analytically~\cite{gunnwarner}, though here we choose not to do so for simplicity. For even higher momentum, the central value of the Gaussian continues to shift to higher energies, as is apparent from the bottom right panel in Fig.~\ref{fig:freeion}. From \eqref{eq:sko_impulseapprox} one can furthermore see that the structure factor asymptotes to $\delta$-function in the $q\to \infty$ limit, correctly reproducing the free particle limit.

It then remains to be shown that the asymptotic form in \eqref{eq:sko_impulseapprox} is equivalent to using the plane wave approximation for the recoiling nucleus. We start from the position space wave function of a nucleus in an isotropic harmonic potential with characteristic frequency $\omega'$ 
\begin{equation}
\langle\bfr_N \ket{\psi_0}=\frac{(\pi m_N^3 \omega'^3)^{1/4}}{\pi} e^{-\frac{1}{2}r_N^2 m_N \omega'}
\end{equation}
The ground state wave function of a harmonic crystal is the direct product of the ground states of a set of harmonic oscillator, weighted by the density of states of the crystal 
\begin{align}
\langle\bfr_N \ket{\psi_0}&\sim \Pi_{D(\omega')} e^{-\frac{1}{2}r_N^2 m_N \omega'}\\
&= \frac{(\pi m_N^3 \bar\omega^3)^{1/4}}{\pi} e^{-\frac{1}{2}r_N^2 m_N  \bar\omega}.
\end{align}
where we are now working in the $T\approx0$ limit.
In the second line we used the definition \eqref{eq:averagedomega} and normalized the wave function such that $\langle\psi_0\ket{\psi_0}=1$.

Now returning to the definition in \eqref{skodef}, we insert a set of plane waves $\ket{\bfq_N}$ for the final states 
\begin{align}
S(\bfq,\omega)  =&V\!\int\!\frac{d^3\bfq_N}{(2\pi)^3} \left|\bra{\bfq_N}e^{i \bfq\cdot \bfr_N} \ket{\psi_0}\right|^2 \delta \left(\frac{q_N^2}{2m_N}-E_0 -\omega\right)
\end{align}
with $E_0$ the ground state energy. Plugging in the plane wave form for $\ket{\bfq_N}$ and shifting the integration variable $\bfq_N \to \bfq_N + \bfq$, this yields
\begin{widetext}
\begin{align}
S(\bfq,\omega)  =&\frac{(\pi m_N^3 \bar\omega^3)^{1/2}}{\pi^2} \int\!\frac{d^3\bfq_N}{(2\pi)^3} \left|\int\!d^3\bfr_N e^{-\frac{1}{2}r_N^2 m_N \bar \omega-i \bfq_N \cdot \bfr_N} \right|^2 \delta \left(\frac{q_N^2+ 2\bfq\cdot\bfq_N+q^2}{2m_N}-E_0 -\omega\right)\\
=&\frac{8 \pi^{3/2}}{(m_N\bar\omega)^{3/2}} \int\!\frac{d^3\bfq_N}{(2\pi)^3}  e^{-\frac{q_N^2}{m_N \bar\omega}}  \delta \left(\frac{q_N^2+ 2\bfq\cdot\bfq_N+q^2}{2m_N}-E_0 -\omega\right).
\end{align}
\end{widetext}
The impulse approximation is a good approximation when $q\gg\sqrt{m_N \bar\omega}$, while the integrand is exponentially suppressed unless $q_N\lesssim\sqrt{m_N \bar\omega}$. We can therefore drop the $q_N^2/2m_N$ term from the $\delta$-function, as well as the ground state energy $E_0$. Carrying out the residual integrals then delivers the same expression as in \eqref{eq:sko_impulseapprox}. We therefore conclude that at low $T$, the usage of plane wave wavefunctions for the excited states for the recoiling nucleus is justified in the limit $q\gg\sqrt{m_N \bar\omega}$, as long as the bound state nature of the initial state is accounted for.

\section{Details of rate calculation\label{app:rates}}

In this appendix, we provide expressions for the Migdal rate, $dR/d\omega$, in the free-particle and impulse approximations in the soft limit. We also illustrate the theoretical uncertainties arising from our various approximations. The differential rate is given by $dR/d\omega = N_T n_\chi \int d^3 \bfv_\chi \, f(\bfv_\chi) \, v_\chi \, d\sigma/d\omega$, where $N_T$ is number of target nuclei per kilogram and $n_\chi = \rho_\chi/m_\chi$ is the DM number density. Consider first the free particle approximation in the soft limit. The differential cross-section entering the rate is then given by \eqref{eq:freeionfinal}. We approximate the ELF by defining Im$[-\epsilon(k',\omega)] \equiv \text{Im}[-\overline{\epsilon_{\bfK\bfK}(\bfk,\omega)}]$ where $k' \equiv |\bfK+\bfk|$ can lie outside the 1BZ and the overline corresponds to the angular average for fixed $k'$ and interpolation between the discrete $k'$ mesh points sampled by the DFT code \texttt{GPAW}. (We neglect the off-diagonal parts of the dielectric matrix.) The $d^3\bfk$ integral can then be extended outside the 1BZ. In the soft limit, the $\bfk$-dependence can be factorized and isolated into the quantity
\begin{equation}
\begin{aligned}
I(\omega)&\equiv \frac{8\alpha}{3 (2\pi)^2 \omega^4} \int \!\! dk \, k^2 \,\Zion^2(k)\, \text{Im}\left[\frac{-1}{\epsilon(k,\omega)}\right].
\end{aligned}
\end{equation}
Due to this factorization, we can use the delta function to perform the angular $\bfq_N$ integral. We then convert the $q_N$ integral to an integral over $E_N$, the nuclear recoil energy, which can also be performed straightforwardly. The result for the differential rate becomes
\begin{equation}
\begin{aligned}
\frac{dR^{\rm free}}{d\omega} \approx &\frac{N_T \rho_\chi}{m_\chi} A^2 \sigma_n  \int_{v_{\rm min}}^{v_{\rm max}} \frac{d^3{\bfv}}{v} \frac{f(\bfv)}{2\mu_{\chi n}^2}  \\
&\times \Theta\left(v_{\rm max}-v_{\rm min}\right) \left( E_{N,\,\rm max }^2 -  E_{N,\,\rm min}^2 \right)  \\
&\times \Theta\left(E_{N,\,\rm max }-E_{N,\,\rm min }\right) I(\omega)
\end{aligned}
\end{equation}
where $f(\bfv)$ is the standard halo model for the DM velocity distribution in the Earth frame, and $v_{\rm min} = \sqrt{\frac{2\omega}{\mu_{\chi N}}}$,  $v_{\rm max} = v_{esc}+v_e$ are limits on the $|\bfv|$ integration. We have also defined the minimum and maximum nuclear recoil energies
\begin{equation}
\begin{aligned}
E_{N,\,\rm min}\equiv &\text{max}\Big[E_N^{\rm th}, \frac{\mu _{\chi N}}{m_N} \big(-v \sqrt{\mu _{\chi N} (v^2 \mu
   _{\chi N}-2 \omega )}   \\
   &   + v^2 \mu _{\chi N}-\omega\big)\Big]\\
E_{N,\,\rm max} \equiv &\frac{\mu _{\chi N} \left(v \sqrt{\mu _{\chi N} \left(v^2 \mu
   _{\chi N}-2 \omega \right)}+v^2 \mu _{\chi N}-\omega \right)}{m_N}
\end{aligned}
\end{equation}
where $E_N^{\rm th}$ cuts off the $E_N$ integration to account for a possible energy threshold. $\mu_{\chi N}$ is the DM--nucleus reduced mass. The DM velocity integral is straightforward to perform numerically. 

\begin{figure*}[!t]
\includegraphics[width=0.48\textwidth]{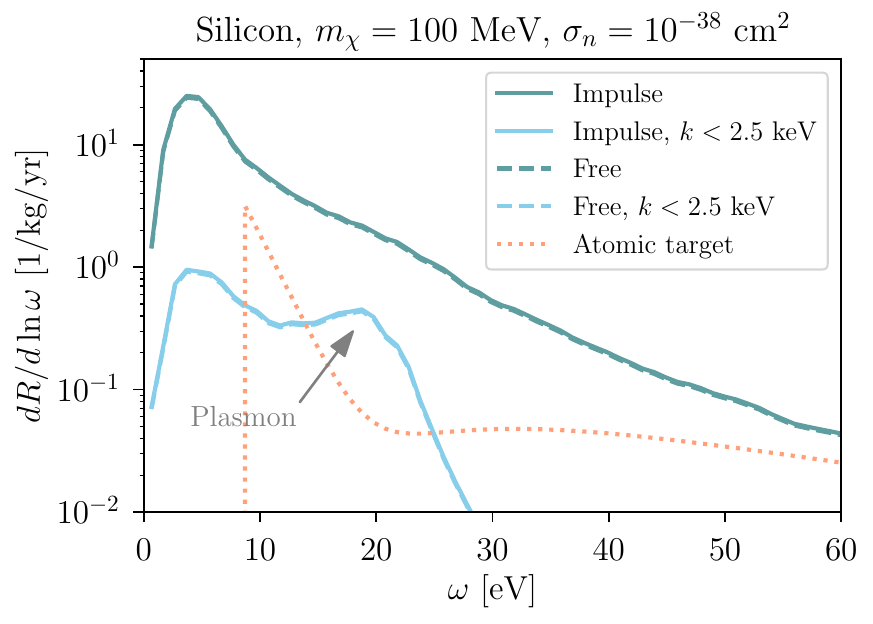}\quad \includegraphics[width=0.48\textwidth]{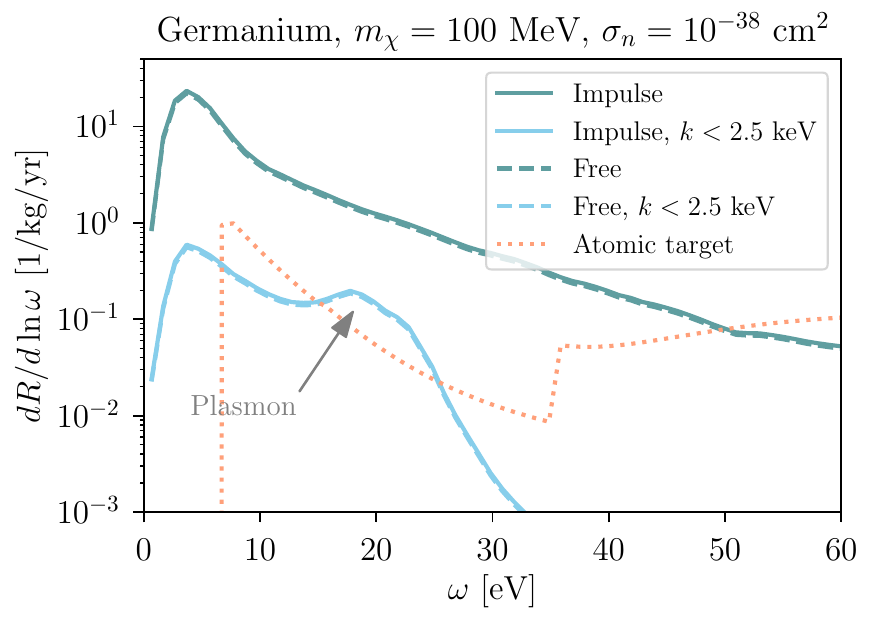}
\caption{Differential Migdal ionization rate in Si (left) and Ge (right) for $m_{\chi}=100$ MeV and $\sigma_n=10^{-38}$ cm$^2$. The total rate in the impulse (free particle) approximation with $E_N>4\bar\omega$ is given by the solid (dashed) curves. The corresponding result cutting off the $k$ integral at 2.5 keV is shown in light blue, where the plasmon resonance (which is only present for small $k$) is clearly visible. For comparison, the dotted curves show the atomic  ionization rate due to the Migdal effect, computed in Ref.~\cite{Ibe:2017yqa} for Si and Ge atoms.    \label{fig:rates}}
\end{figure*}

The corresponding derivation in the impulse approximation proceeds similarly, except we use \eqref{eq:ratefinalsoft} for $d\sigma/d\omega$. In the soft limit, the $\bfk$ integral again factorizes into $I(\omega)$ given above. There is an additional integral over $\bfp_f$ in this case due to the appearance of the form factor $F$ rather than the usual momentum delta function. We shift the $\bfp_f$ integration to be over $\bfq = \bfp_i-\bfp_f$. We then first perform the angular $\bfq_N$ integral, and use the energy delta function to perform the angular $\bfq$ integral (the azimuthal integrals are trivial). The $|\bfq|$ integration can then be performed, followed by the integral over $q_N$, which we trade for an integral over $E_N$. In doing so, we again cut off the $E_N$ integration at a threshold value $E_N^{\rm th}$, which we will vary to illustrate the theoretical uncertainties associated with the impulse approximation. We could have instead swapped the order of $q$ and $q_N$ integration and imposed a cutoff on $q$, but we find that the two procedures yield similar results for $m_{\chi} \gtrsim 50$ MeV. The DM velocity integral proceeds as before, only with $v_{\rm min}= \sqrt{2(\omega + E_N^{\rm th})/m_{\chi}}$. The result for the rate is then
\begin{equation}
\begin{aligned}
\frac{dR}{d\omega} \approx &\frac{N_T \rho_\chi}{m_\chi} A^2 \sigma_n  \int_{v_{\rm min}}^{v_{\rm max}} \frac{d^3\bf{v}}{v} \frac{f(\bfv)}{\mu_{\chi n}^2} \\
&\times \Theta\left(v_{\rm max}-v_{\rm min}\right) G^{\rm IA}(\omega,v) I(\omega)
\end{aligned}
\end{equation}
where
\begin{equation}
\begin{aligned}
&G^{\rm IA}(\omega,v) \equiv  \int^{\frac{ m_\chi v^2}{2} -\omega}_{E_N^{\rm th}}\!\frac{dE_R E_R }{2}  \Theta\left(\frac{ m_{\chi}v^2}{2}-\omega -E_N^{\rm th}  \right)  \\
&\left[  \text{Erf}\left( \frac{q_{\rm max}-\sqrt{2 m_N E_N}}{\sqrt{m_N \bar \omega}}\right) + \text{Erf}\left( \frac{q_{\rm min}+\sqrt{2 m_N E_N}}{\sqrt{m_N \bar \omega}}\right) \right. \\
&\left.  - \text{Erf}\left( \frac{q_{\rm min}-\sqrt{2 m_N E_N}}{\sqrt{m_N \bar \omega}}\right) -\text{Erf}\left( \frac{q_{\rm max}+\sqrt{2 m_N E_N}}{\sqrt{m_N \bar \omega}}\right)  \right]
\end{aligned}
\end{equation}
and
\begin{equation}
\begin{aligned}
q_{\rm min}&\equiv m_\chi v - \sqrt{2m_\chi\left(\frac{m_\chi v^2}{2} - E_N - \omega \right) }\\
q_{\rm max}&\equiv m_\chi v + \sqrt{2m_\chi\left(\frac{m_\chi v^2}{2} - E_N - \omega \right) } \label{eq:q_lims}.
\end{aligned}
\end{equation}

In Fig.~\ref{fig:freeion} we showed that the impulse approximation starts to break down for \mbox{$q/\sqrt{2\bar\omega m_N}\lesssim3$}. To estimate the size of the uncertainty associated with this approximation, we calculate the rate for both \mbox{$E_N^{\rm th}=4\bar\omega$} and \mbox{$E_N^{\rm th}=9\bar\omega$}, which is equivalent to restricting the phase space to respectively \mbox{$q/\sqrt{2\bar\omega m_N}>2$} and \mbox{$q/\sqrt{2\bar\omega m_N}>3$} in the free particle limit. The difference is shown by the shaded bands in Fig.~\ref{fig:sensitivity}. For $m_\chi\lesssim 50$ MeV, the rate is dominated by the phase space corresponding to
\mbox{$E_N^{\rm th}<9\bar\omega$},
as is evident by the diverging uncertainty bands. Here the impulse approximation ceases to be reliable and we chose not to extrapolate our results to this regime.

In Fig.~\ref{fig:rates} we show the differential rate for $m_\chi=100$ MeV for Si and Ge under different cuts and assumptions. First, one observes that the free ion and impulse approximations are essentially identical for this mass point. To isolate the contribution from the plasmon pole in particular, we also plot the rate with a $k<2.5$ keV cut, which reveals the plasmon enhancement around $\omega\approx 20$ eV. The separation between the blue and the green curves however shows that this contribution is highly subleading as compared to the high $k$, off-resonance part of the ELF. Finally, in orange we show the rate for the Migdal effect in atomic Si or Ge, using the results of \cite{Ibe:2017yqa}; this case corresponds to the collision of the DM with a single, isolated Si or Ge atom, and includes only atomic ionizations as a possible charge signal. Clearly, this gives a poor approximation for the rate in semiconductors, which is much larger due to the reduced energies for valence states, as well as the possibility of detecting all electronic excitations between bands. In contrast, for the atomic Migdal effect, transitions between bound electronic states do not give an ionization signal.

\begin{figure}[!t]
\includegraphics[width=0.48\textwidth]{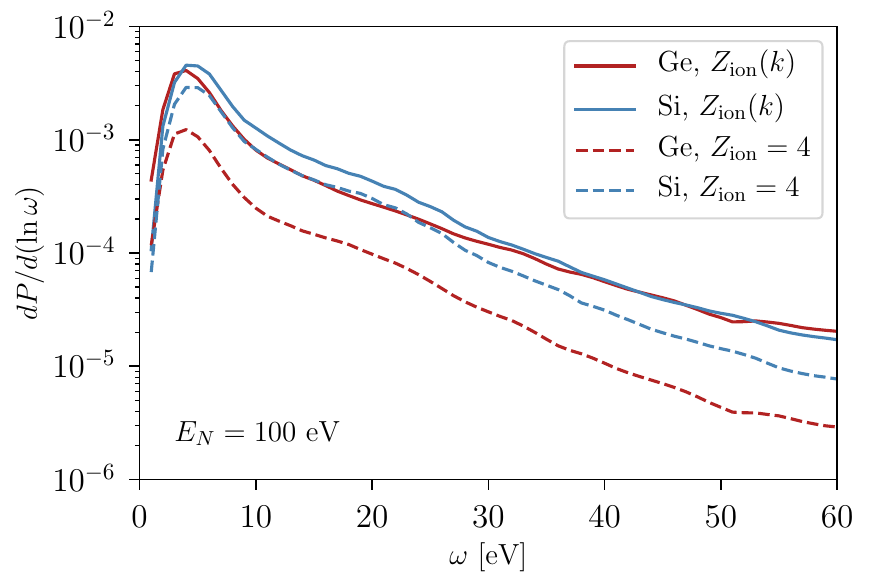}
\caption{Differential Migdal ionization probability $dP/d \ln \omega$ in Si and Ge for fixed nuclear recoil energy $E_N=100$ eV. The solid lines are our default calculation with momentum dependent $\Zion(k)$, while the dashed lines are for fixed $Z_{\rm ion} = 4$. \label{fig:rateZion}}
\end{figure}

In Fig.~\ref{fig:rateZion}, we show the impact of using momentum-dependent $\Zion(k)$ (the default for results in this paper) as opposed to constant $\Zion$. The momentum-dependent ion charges were obtained using the results of Ref.~\cite{BrownXray}, which provides tabulated form factors for Si$^{4+}$ and Ge$^{4+}$ as a function of $k/(4\pi)$ in \AA$^{-1}$, where $Z_{\rm ion}(k)$ starts deviating from $Z_{\rm ion} = 4$ at $k \sim 5-10$ keV.  The rate with momentum-dependent charges is larger by a factor of $3$ for Ge and 1.6 for Si, where the effect is larger for Ge because the semi-core electrons are not as tightly bound. The difference with constant $\Zion$ also becomes larger at higher $\omega$, where the differential rate is also weighted at higher momentum transfers.

\section{Semiclassical derivation of atomic Migdal effect} \label{sec:semiclassical_atomic}

In the main text, we interpreted the Migdal effect as electronic transitions due to the potential of a recoiling nucleus. Here we provide a semiclassical derivation of the Migdal effect in atoms that reinforces this interpretation by remaining in the lab frame. A similar approach has been used in calculating the Migdal effect for nuclear decay processes~\cite{LAW1977339,PhysRev.90.11}.

In the semiclassical approach, the nucleus motion is treated classically and the electron degrees of freedom only depend parametrically on the nucleus positions (also known as the Born-Oppenheimer approximation). Suppose the nucleus is initially at the origin, and gets an impulse to velocity $\bfv_N$ at $t=0$. This is often also called an impulse approximation; however, it implies a stronger set of conditions than the impulse approximation used for our semiconductor calculation, since it treats the nucleus classically and with an instantaneous change to its velocity. The electron Hamiltonian can then be written as
\begin{align}
	H(t) = H_{0}  +H_1(t)
\end{align}
where $H_0$ is the time-independent (or equilibrium) Hamiltonian of the electronic system in the presence of a nucleus at the origin. $H_1(t)$ is the time dependent perturbation which describes the effect of the recoiling nucleus with charge $Z_N$:
\begin{align}\label{eq:H1}
	H_1(t) = -  \sum_{\beta} \frac{Z_{N} \alpha}{| \bfr_{\beta} - {\bf R}_{N}(t)|} +  \sum_{\beta} \frac{Z_{N} \alpha}{| \bfr_{\beta}|}
\end{align}
with the nucleus position ${\bf R}_{N}(t) = \theta(t) \bfv_N t$. Here we summed over all electrons $\beta$. The second term in $H_1(t)$ is important to subtract the equilibrium contribution of the nucleus to $H_0$, since for $t>0$ the nucleus is no longer in its equilibrium position. Thus, it ensures that $H_1$ is switched off for $t<0$. The time scale for the dipole transitions will be $t \simeq 1/(\alpha_{em}^2 m_e)$, which satisfies $v_N t \ll r_{\rm Bohr} \ll ct$ for sub-GeV dark matter scattering. The first condition $v_N t \ll r_{\rm Bohr}$ justifies the expansion in small ${\bf R}_N$ compared to the typical extent of electronic wavefunctions, while the second condition $r_{\rm Bohr} \ll t$ justifies the use of the electrostatic potential.

For small $t$, we can expand the potential about small ${\bf R}_{N}(t)$:
\begin{align}
	H_1(t) \approx - \sum_{\beta} \frac{Z_{N} \alpha \hat \bfr_{\beta} \cdot {\bf R}_{N}(t)}{\bfr_{\beta}^{2}} 
\end{align}
which is just the dipole potential for a recoiling nucleus with dipole moment $Z_{N} {\bf R}_N(t)$. In time-dependent perturbation theory, we can now compute the transition probability as
\begin{align}
    P_{i \to f} &= \left| \frac{1}{\omega} \int_0^\infty dt\, e^{i (\omega + i \eta) t}  \langle f | \frac{dH_1(t)}{dt} | i \rangle \right|^2 \\
    & = \Big| \langle f | \frac{1}{\omega^2}\sum_{\beta} \frac{Z_{N} \alpha \hat \bfr_{\beta} \cdot \bfv_N}{\bfr_{\beta}^{2}} | i \rangle \Big|^2
\end{align}
where $\omega = E_f - E_i$ and $\eta$ is a small positive number that accounts for dissipation and turns off the potential at large times. In the last line, we took $\eta \to 0$. This is precisely the matrix element we obtained in the main text by rewriting \eqref{eq:M_boost} in terms of \eqref{eq:dipoleforce2} with operator identities. 

This gives a physical interpretation for the Migdal effect that is more difficult to see from the boosting argument. To further make contact with our semiconductor calculation, note that we can rewrite
\begin{align}
	H_1(t) = \sum_{\beta} 4\pi Z_{N} \alpha \int \frac{d^{3} \bfk}{(2\pi)^{3}} \frac{e^{i\bfk \cdot \bfr_{\beta}} - e^{i \bfk \cdot (\bfr_{\beta} - {\bf R}_N(t))} }{k^2}. \nonumber
\end{align}
Then the transition probability $P_{i \to f}$  for single ionizations is given by
\begin{align}
	\left| \int \frac{d^{3} \bfk}{(2\pi)^{3}}  \langle f | e^{i\bfk \cdot \bfr} | i \rangle \frac{ 4\pi  Z_{N} \alpha}{k^2}\left( \frac{1}{\omega - \bfv_{N} \cdot \bfk} - \frac{1}{\omega} \right) \right|^2,
\end{align}
and we see the same form of the matrix element as in the semiconductor calculation, \eqref{eq:Mfinal}, up to overall factors that account for the differences in the physical systems and for the DM--nucleus scattering rate. The situation is analogous to bremsstrahlung, which can be computed either in a semi-classical approximation (as in this section) or with quantum mechanics (as we did in the semiconductor case, in Appendix~\ref{app:semi_fullderivation}). Summing over initial and final states and taking the soft limit $\bfv_N \cdot \bfk \ll \omega$ of this transition probability, we can then obtain \eqref{eq:atomic_fourier}.

\bibliography{dielectric}

\end{document}